\documentclass[epj,nopacs,fleqn]{svjour}
\usepackage{graphicx,amsmath,amssymb,slashed,cite,verbatim,multirow,url}
\usepackage{color}
\usepackage{color}
\usepackage{hep}
\usepackage{booktabs}

\usepackage{array}
\newcolumntype{?}{!{\vrule width 1pt}}

\newcommand{\mgamc}{{\sc MG5aMC}}

\begin{document}


\title{Dark matter production through loop-induced processes at the LHC: the s-channel mediator case}

\author{
Olivier Mattelaer\inst{1},
Eleni Vryonidou\inst{2}}

\institute{ 
Institute for Particle Physics Phenomenology (IPPP), Durham University, Durham, DH1 3LF,\\
United Kingdom
 \and  Centre for Cosmology, Particle Physics and Phenomenology (CP3),
 Universit\'e catholique de Louvain,\\
 B-1348 Louvain-la-Neuve, Belgium
 }
\date{}
\abstract
{We show how studies relevant for mono-X searches at the LHC in simplified models featuring a dark matter candidate and an $s$-channel mediator can be performed within the {\sc MadGraph5\_aMC@NLO} framework. We focus on gluon-initiated loop-induced processes, mostly relevant to the case where the mediator couples preferentially to third generation quarks and in particular to the top quark. Our implementation allows us to study signatures at hadron colliders involving missing transverse energy plus jets or plus neutral bosons ($\gamma,Z,H$), possibly including the effects of extra radiation by multi-parton merging and matching to the parton shower. }

\titlerunning{Dark matter production through loop-induced processes at the LHC}   
\authorrunning{Mattelaer and Vryonidou}
\maketitle

\vspace*{-11.5cm}
{\noindent\small{IPPP/15/48, DCPT/15/96, CP3-15-23, MCnet-15-20}}
\vspace*{9.7cm}

\section{Introduction}
\label{sec:intro}
Searching for Dark Matter (DM) candidates is a leading priority of Run II at the LHC. As the DM particles are expected to be stable and neutral under the Standard Model (SM) interactions,  
they do not leave any trace in the detectors. Therefore, the DM searches focus on the associated production with visible SM particles. These processes lead to missing transverse energy and are often classified as mono-X searches, where X can be jet(s), photon, $Z$, $W$ or even a Higgs boson.

DM model building is a very rich and exciting theoretical activity, with many mechanisms and models being explored in the community. Such a wealth of options, however, makes top-down searches somewhat limited in scope and calls for more general bottom-up approaches. These are generally classified into two groups.  The first involves effective operators, describing the direct interaction between the DM particles and the SM ones, via higher dimensional operators.  These so-called Effective Field Theory (EFT) models arise in the presence of sufficiently heavy additional states, not directly accessible in our experiments,  that can be integrated out leading to a contact interaction between the dark matter and the SM particles. In this case, experimental searches can set limits on the scale $\Lambda$ (assuming unit coupling) which determines the strength of the interaction. 

The second class of models, called simplified models of DM, is based on a single particle exchange between the DM and SM particles. The particle exchanged is called the mediator,  a neutral boson in the case of  $s$-channel mediator, or a more generic state in the case of $t$-channel ones. Simplified models generally include more parameters than EFT models, as the couplings of the mediator to the DM and SM particles, the width and mass of the mediator play a role in the phenomenology, and hence they can parametrise more reliably the kinematics of the production process when the mediator is not sufficiently heavy to be integrated out.

Both classes have been extensively studied in the literature.   Simplified models for DM are presented for example in \cite{Carpenter:2013xra},  while a review of simplified DM models and their relevance for missing energy searches at the LHC is given in \cite{Abdallah:2014hon}.
The EFT limit of these interactions for the scalar mediator case was studied recently in \cite{Haisch:2015ioa}, where limits on the EFT operators are set, based on the missing transverse energy searches at the LHC. Various other works focus on DM production in the EFT, providing results at NLO in QCD for various mono-X processes \cite{Mao:2014rga,Fox:2012ru,Wang:2011sx,Huang:2012hs}. 
We refer the interested reader to the recent reviews in Refs.~\cite{Abdallah:2015ter,Abercrombie:2015wmb} for a more complete list of models and processes to be studied at the LHC.

Within both classes of models, an especially interesting scenario arises when either the dark matter or the mediator only couples  to the third generation quarks and in particular to the top quark. In this case, the only DM production mechanism at tree-level becomes production in association with heavy quarks, while all other processes have to proceed through heavy quark loops in gluon fusion. Due to the high gluon luminosity at the LHC, these processes can be important and have already received attention in the literature. Within the EFT models, this scenario has been studied in \cite{Frandsen:2012db,Haisch:2015ioa} for the mono-jet signal for scalar or pseudoscalar contact interactions. In the same EFT setup, \cite{Haisch:2013fla} considers the production of DM in association with two jets, whose angular correlations can provide a handle to distinguish between scalar and pseudoscalar interactions. In the context of a simplified model,  the mono-jet process from top quark loops for a scalar and pseudoscalar mediator has also been discussed \cite{Buckley:2014fba,Harris:2014hga}.

In this work, we present results of the implementation of a simplified DM model in the {\sc MadGraph5\_aMC@NLO} framework (\mgamc\ henceforth) \cite{Alwall:2014hca}, 
including a Dirac fermion as dark matter and either a scalar, pseudoscalar, vector or  axial-vector mediator. In particular, we focus on the case where the mediator couples only to the top quark (and to bottom in the case of an axial vector to avoid anomalies), and study the loop-induced contributions to the most commonly searched for  mono-X processes for various masses of the mediator and the dark matter particles.

This paper is organised as follows. In section \ref{model} we briefly describe the simplified model employed, and in section \ref{setup} the calculation setup. In section \ref{jets} we present results for the jets plus missing transverse energy signature comparing  the full top-quark mass dependence with the top-EFT approach while in section \ref{monoX} we discuss mono-Z, Higgs and photon processes. We  draw our conclusions in section \ref{conclusions}.

\section{Simplified Model}
\label{model}

We assume a simplified model where the mediator only couples to top quarks and to the DM particles.
The DM particle ($\chi$) is taken to be a Dirac fermion in this study but the implementation of the model (see next Section) is flexible to allow a real or complex scalar.
The mediator can be chosen to be either a scalar ($Y_0$) or a vector ($Y_1$). In the scalar case,  the interaction Lagrangian is given by:
\begin{eqnarray}\nonumber
\mathcal{L}_{DM}^{Y_0}&=&\bar{\chi} (g^S_{DM}+i g^P_{DM}  \gamma^5)\chi \, Y_0, \\
\mathcal{L}_{SM}^{Y_0}&=&\bar{t} \frac{y^t}{\sqrt{2}}(g^S_{t}+ig^P_{t} \gamma^5) t \,Y_0.
\label{langX0}
\end{eqnarray}
For convenience we normalise the scalar and pseudoscalar interaction with the top quark ($t$) by the top Yukawa coupling ($y^t = \frac{m_t}{v}$).
In the vector case,  the interaction Lagrangian is given by:
\begin{eqnarray}\nonumber
\mathcal{L}_{DM}^{Y_1}&=&\bar{\chi} \gamma_{\mu} (g^V_{DM}+g^A_{DM}  \gamma^5)  \chi \, Y_1^{\mu}, \\
\mathcal{L}_{SM}^{Y_1}&=& \bar{t}\gamma_{\mu}(g^V_{t}+g^A_{t} \gamma^5)  t \, Y_1^{\mu} + \bar{b}\gamma_{\mu}(-g^A_{t} \gamma^5)  b \, Y_1^{\mu}.
\label{langX1}
\end{eqnarray} 
For the axial-vector case, we also introduce a coupling to bottom quarks ($b$) and fixed it to be opposite to the one of the top quark ($g^{A}_t$) to ensure the cancellation of the gauge anomaly. 

The couplings, mass and width of the mediator and the mass of the DM can be extracted from the predictions of more concrete DM models. In particular, the width of the mediator can be computed by considering the particle content of $\mathcal{L}_{SM}+\mathcal{L}_{DM}$ and possibly of a more involved dark sector.

\section{Method and tools}
\label{setup}
\subsection{Technical setup}
The computation is performed within the \mgamc\ framework. For this study we employ the latest version (2.3.0), which allows automatic event generation for loop induced processes  \cite{Hirschi:2015iia}.  Within \mgamc, {\sc MadLoop}~\cite{Hirschi:2011pa}  computes the one--loop amplitudes, using the {\sc OPP} integrand reduction method~\cite{Ossola:2006us} (as implemented in {\sc CutTools}~\cite{Ossola:2007ax}).
We use the UFO model of \cite{Mawatari}, generated by  {\sc FeynRules}/ {\sc NLOCT} \cite{Alloul:2013bka,Degrande:2011ua}.
The model can be downloaded from the {\sc FeynRules} catalogue of models \cite{feynrules}.
 We stress here that the implementation of the model has been validated by comparing to the SM. This was achieved by setting various parameters to the corresponding SM values and studying processes with a Higgs and a $Z$, mimicked by a $Y_0$ and $Y_1$ respectively, and perfect agreement has been found. 

As the processes we consider in this work start at one loop (at leading order), NLO corrections are  not available.
Such NLO computation would require some challenging two-loop multi-scale integrals, most of which are not yet available.
In order to provide a better description of the kinematics, one can employ the method of Matrix-Element--Parton Shower (ME+PS) matching/merging \cite{Alwall:2007fs}.  ME+PS  schemes allow the consistent combination of matrix elements with different jet multiplicities via their matching to a parton shower. Merging of samples of different multiplicities in loop-induced processes has been done within \mgamc\  in \cite{Alwall:2011cy} for H+jets and more recently in \cite{Hespel:2015zea} for $ZH$ associated production, by employing a reweighting procedure. 
ME+PS results for Higgs plus jets, obtained automatically with the interface of  \mgamc\   to {\sc Pythia6} \cite{Sjostrand:2006za} have been presented in \cite{Hirschi:2015iia}.

The implementation of ME+PS method in \mgamc\ comes in two variants: the traditional $k_T$-MLM and the shower-$k_T$ schemes. The two give comparable results as discussed in \cite{Alwall:2008qv}. In this study we will employ the traditional $k_T$-MLM  scheme, and in particular the most recent implementation of the scheme in conjunction with {\sc Pythia8}  \cite{Sjostrand:2007gs,Sjostrand:2014zea}, for events generated in \mgamc. 

The value of the merging scale $Q_{\rm{cut}}$ is selected on a process-by-process basis to ensure that there is a smooth transition between the ME and PS regimes. In practice, this is assessed by examining the differential jet rate distributions, which show whether the transition is indeed smooth. The differential jet rates are obtained by customising the {\sc Pythia8} routines. 
The other distributions are obtained by passing the merged samples through {\sc MadAnalysis5} \cite{Conte:2012fm, Conte:2014zja}, which is interfaced to {\sc FastJet} \cite{Cacciari:2011ma} for jet reconstruction. For the jet clustering we use a minimum jet transverse momentum of 25 GeV, and employ the anti-$k_T$ algorithm \cite{Cacciari:2008gp} with a radius of $R=0.4$.

While it is straightforward to study ME+PS merging for all mono-X processes considered here, for the sake of brevity and simplicity we will only present merged sample results for the jets and missing transverse energy signal, while for the rest of the processes we will only show parton-level results.

For the results presented here,  we use the MSTW2008LO \cite{Martin:2009iq} parton distribution functions (PDFs) and the central renormalisation and factorisation scales are set to half the sum of the transverse masses of the final state particles: $\mu^0=\mu^0_R= \mu^0_F=\frac{1}{2}\sum_i \sqrt{m_i^2+p_{T,i}^2} $. In our results, scale variations are obtained by varying the scales independently in the range of $\mu^0/2< \mu_{R,F} < 2\mu^0 $, computed automatically with {\sc SysCalc} \cite{SysCalc}. All results will be presented for the LHC with 13 TeV centre-of-mass energy. The computation is performed within the 5-flavour scheme, unless otherwise stated. 

\subsection{Benchmark points}

In order to demonstrate various physics effects within this model, we concentrate on a series of benchmarks.
In all cases, we assume the same nature of coupling between mediator-DM and mediator-tops and concentrate on four scenarios: the scalar one ($g^{S}_{DM} = g^{S}_t=1$) ,
the pseudo-scalar one ($g^{P}_{DM} = g^{P}_t=1$), the vector one ($g^{V}_{DM} = g^{V}_t=1$) and the axial-vector one ($g^{A}_{DM} = g^{A}_t=1$). The non-specified couplings in these four scenarios are set to zero. 
We also select three mass benchmarks, shown in Table \ref{tab:points}.  We find this choice illustrative as these represent three different physics cases.
The first benchmark dubbed `\emph{resonant}' corresponds to the resonant production of the mediator and its decay into a pair of dark-matter particles. The second benchmark `\emph{heavy mediator}' involves a heavy mediator and a light dark matter particle. The width associated to the mediator is, in that case, quite large and interesting off-shell effects can be observed. The final benchmark `\emph{heavy DM}' corresponds to the case where the mediator can not be produced on-shell, as its mass lies below the $2m_{\chi}$ threshold.

\begin{table*}[t]
\renewcommand{\arraystretch}{1.3}
\begin{center}
    \begin{tabular}{l? c | c |c }
        \hline \hline
      Benchmark &  Resonant & Heavy mediator & Heavy DM\\ \specialrule{1pt}{0pt}{0pt}
    Mediator mass   & 200 &1000  &  400\\         \hline
     Dark matter mass &  50 & 1 & 500 \\   \hline \hline
\end{tabular}
 \caption{\label{tab:points} Mass benchmarks in GeV.}  
\end{center} 
\end{table*}

\begin{table*}[t]
\renewcommand{\arraystretch}{1.3}
\begin{center}
    \begin{tabular}{l ? c | c | c | c}
        \hline \hline
  Benchmark & S & P & V & A \\ \specialrule{1pt}{0pt}{0pt}
  Resonant  & 5.17  & 6.89  & 5.17 & 19.3 \\         \hline
  Heavy Mediator & 88.0  & 94.5 & 105.7 & 172\\ \hline
  Heavy DM &  3.10  & 11.89  & 22.2  & 36.0\\   \hline \hline
\end{tabular}
 \caption{\label{tab:width} LO widths in GeV for the various mediators. The computation of the width has been performed with {\tt MadWidth} \cite{Alwall:2014bza}.}  
\end{center} 
\end{table*}

For completeness, we show the mediator widths obtained for the three benchmark points and  for the four possible couplings scalar (S), pseudoscalar (P), vector (V) and axial-vector (A) in Table \ref{tab:width}. In the first scenario the mediator decays only to DM, in the second to DM and top quarks and in the third to top quarks only. For the axial vector mediator decay into bottom quarks is allowed and included in the width computation for all benchmarks. The computation of the width has been performed at tree-level with {\tt MadWidth} \cite{Alwall:2014bza}.

\section{Jets and missing transverse energy signal} 
\label{jets}
\subsection{Total cross-section results for the mono-jet process}
\label{subsec_xsec}
\begin{figure*}[t]
\centering
\includegraphics[scale=0.5]{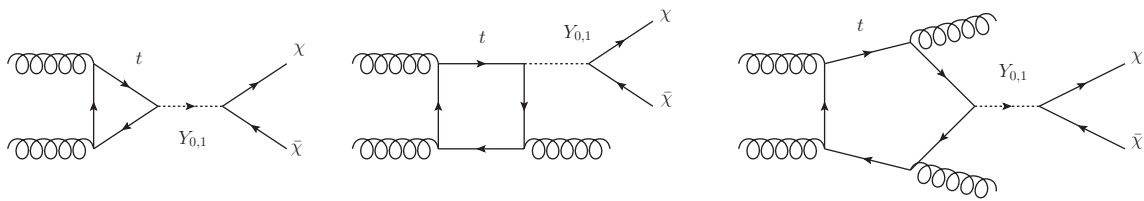}
\caption{Feynman diagrams contributing to jets plus missing transverse energy signal in the simplified model.}
\label{monojet}
\end{figure*}

The first process we consider is the production in association with QCD jets, which leads to a missing energy signature though the production of the mediator and its decay to DM particles. A sample of the Feynman diagrams contributing to this signal are shown in Fig. \ref{monojet}. 

In this study we will be focussing on the shapes of the differential distributions, and therefore we will typically show normalised ones. Nevertheless, we find it instructive to start by presenting the cross-sections we obtain for the mono-jet process for the various benchmarks discussed above, for two cuts on the jet transverse momentum, a minimal cut of 50~GeV and a more realistic one of 200 GeV. In this simple process at the parton-level, the jet transverse momentum is identical to the missing transverse energy, as the DM pair is recoiling against the parton/jet. Results are shown in Table \ref{tab:sigma} for the scalar and pseudoscalar mediators, including the scale and PDF uncertainties for the LHC at 13 TeV.\footnote{We note here that the corresponding LO results for $Z (\to \nu \bar{\nu})+$jet are 711 and 11.7 pb for a cut of 50 and 200 GeV respectively.}
The scale uncertainties can exceed 50\% in some cases, while the PDF ones rise from $\sim $1\% in the resonant case to 2\% in the heavy DM scenario.  
Such large scale uncertainties are similar to those observed for loop-induced processes within the SM \cite{Hirschi:2015iia}. 
The heavy DM scenario gives the smallest cross-sections, as these are suppressed by the DM pair production threshold which lies at 1~TeV, leading to sub-fb cross-sections. The resonant production is on the other hand enhanced, as expected. The gluon fusion amplitudes differ for a scalar and a pseudoscalar mediator, by a factor of $2/3$ in the infinite top mass limit. This manifests also here with the pseudoscalar mediator giving consistently larger cross-sections. We note here that the heavy mediator scenario gives larger cross-sections than the heavy DM one, as a significant fraction of the cross-section comes from the off-shell region, well below the mass of the mediator (1 TeV). These results have been obtained for the couplings set arbitrarily to 1. By modifying these couplings the cross-sections can be enhanced/reduced accordingly. 

\begin{table*}[t]
\renewcommand{\arraystretch}{1.3}
\begin{center}
\begin{tabular}{l ? c | c | c | c}
\hline \hline
Benchmark & \multicolumn{2}{c| }{Scalar} &\multicolumn{2}{c }{ Pseudoscalar}\\ \hline
Jet $p_T$ cut in GeV & 50 & 200 & 50 & 200 \\ \specialrule{1pt}{0pt}{0pt}
Resonant  & 4.11 $\, {}^{+49\%}_{-31\%}\, {}^{+0.8\%}_{-0.9\%}$  & 0.244\, ${}^{+50\%}_{-31\%}\,{}^{+1.1\%}_{-1.1\%} $  & 10.1\, ${}^{+49\%}_{-31\%}\,{}^{+0.8\%}_{-0.9\%}$ & 0.584\, ${}^{+50\%}_{-31\%}\,{}^{+1.1\%}_{-1.1\%}$ \\         \hline
Heavy mediator &  3.22 \cdot $10^{-2} \, {}^{+55\%}_{-33\%}\,{}^{+1.7\%}_{-1.7\%}$ & 4.92\, \cdot $10^{-3} \, {}^{+55\%}_{-33\%}\,{}^{+1.8\%}_{-1.8\%}$ & 4.23 \cdot $10^{-2} \, {}^{+55\%}_{-33\%}\,{}^{+1.6\%}_{-1.6\%}$ & 6.47\, \cdot $10^{-3} \, {}^{+54\%}_{-33\%}\,{}^{+1.8\%}_{-1.8\%}$ \\ \hline
Heavy DM & 4.33 \cdot $10^{-5} \, {}^{+55\%}_{-33\%}\,{}^{+2.2\%}_{-2.1\%}$ & 8.54 \cdot $10^{-6} \, {}^{+55\%}_{-33\%}\,{}^{+2.3\%}_{-2.2\%} $ & 1.73 \cdot $10^{-4} \, {}^{+54\%}_{-33\%}\,{}^{+2.0\%}_{-2.0\%} $ & 3.35\, \cdot $10^{-5} \, {}^{+54\%}_{-33\%}\,{}^{+2.1\%}_{-2.1\%} $ \\   \hline \hline
\end{tabular}
\caption{\label{tab:sigma} Cross-sections in pb for $p p \to \bar{\chi} \chi j$ at 13 TeV for the scalar and pseudoscalar mediators for a cut of 50 and 200 GeV on the jet transverse momentum and the corresponding scale and PDF uncertainties for the LHC at 13 TeV.  }
\end{center} 
\end{table*}

The corresponding results for the mono-jet cross-sections for the vector and axial-vector mediators along with the scale/PDF uncertainties for the three benchmarks are shown in Table \ref{tab:sigmaV}.  Similar observations regarding the scale and PDF uncertainties can be made in this case. The cross-sections for the three benchmarks with the vector mediator follow the same patterns as those for the scalar and pseudoscalar. We notice however that the axial-vector cross-sections, especially for the heavy DM scenario, are significantly larger than the corresponding vector ones. This effect, which will also be evident in the other mono-X processes studied in the next section, originates from the non-conservation of the  axial vector current. The mediator propagator contains a term proportional to $p^{\mu}p^{\nu}/M_{Y_1}^2$ that when contracted with the axial-vector current leads to terms proportional to $m_t m_{\chi}/M_{Y_1}^2$, and leads to an enhancement of the cross section for large DM masses. 

\begin{table*}[t]
\renewcommand{\arraystretch}{1.3}
\begin{center}
    \begin{tabular}{l ? c | c | c | c}
        \hline \hline
Benchmark & \multicolumn{2}{c| }{Vector} &\multicolumn{2}{c }{Axial-vector}\\ \hline
Jet $p_T$ cut in GeV & 50 & 200 & 50 & 200 \\ \specialrule{1pt}{0pt}{0pt}
Resonant  & 0.487 \, ${}^{+51\%}_{-31\%}\,{}^{+1.1\%}_{-1.2\%}$ & 0.104 \, ${}^{+51\%}_{-32\%}\,{}^{+1.4\%}_{-1.4\%}$ & 11.5 \, ${}^{+50\%}_{-31\%}\,{}^{+0.7\%}_{-0.9\%}$  & 1.02 \, ${}^{+50\%}_{-31\%}\,{}^{+1.1\%}_{-1.1\%}$   \\         \hline
Heavy mediator & 2.68 \cdot $10^{-4} \, {}^{+56\%}_{-34\%}\,{}^{+2.1\%}_{-2.1\%}$ & 1.55\, \cdot $10^{-4} \, {}^{+57\%}_{-34\%}\,{}^{+2.4\%}_{-2.4\%}$ & 5.51\, \cdot $10^{-3} \, {}^{+52\%}_{-32\%}\,{}^{+0.9\%}_{-1.0\%}$ & 8.97\, \cdot $10^{-4} \, {}^{+53\%}_{-32\%}\,{}^{+1.4\%}_{-1.4\%}$ \\ \hline
Heavy DM &  1.48 \cdot $10^{-6} \, {}^{+57\%}_{-34\%}\,{}^{+2.9\%}_{-2.8\%}$&  1.09\, \cdot $10^{-6} \, {}^{+57\%}_{-34\%}\,{}^{+3.0\%}_{-2.9\%} $ & 1.28\, \cdot $10^{-3} \, {}^{+54\%}_{-33\%}\,{}^{+2.0\%}_{-2.0\%}$ & 2.50\, \cdot $10^{-4} \, {}^{+54\%}_{-33\%}\,{}^{+2.1\%}_{-2.1\%} $\\   \hline \hline
\end{tabular}
 \caption{\label{tab:sigmaV} Cross-section in pb  for $p p \to \bar{\chi} \chi j$ for the vector and axial-vector mediators for a cut of 50 and 200 GeV on the jet transverse momentum and the corresponding scale and PDF uncertainties  for the LHC at 13 TeV.}  
\end{center} 
\end{table*}

\subsection{Differential results for the scalar and pseudoscalar mediators}
To provide an accurate description of the kinematics beyond LO, we consider merged samples of 0, 1 and 2-jet multiplicities. For comparison purposes, in the plots shown below we also include the main background, $Z(\to \nu \bar{\nu}) + $jets, obtained for consistency with LO merging of 0, 1 and 2-jet samples. Both computations are performed in the setup described in the previous section.  

\begin{figure}[tb]
\begin{minipage}[t]{0.5\linewidth}
\centering
\includegraphics[trim=1cm 0 0 0,scale=0.56]{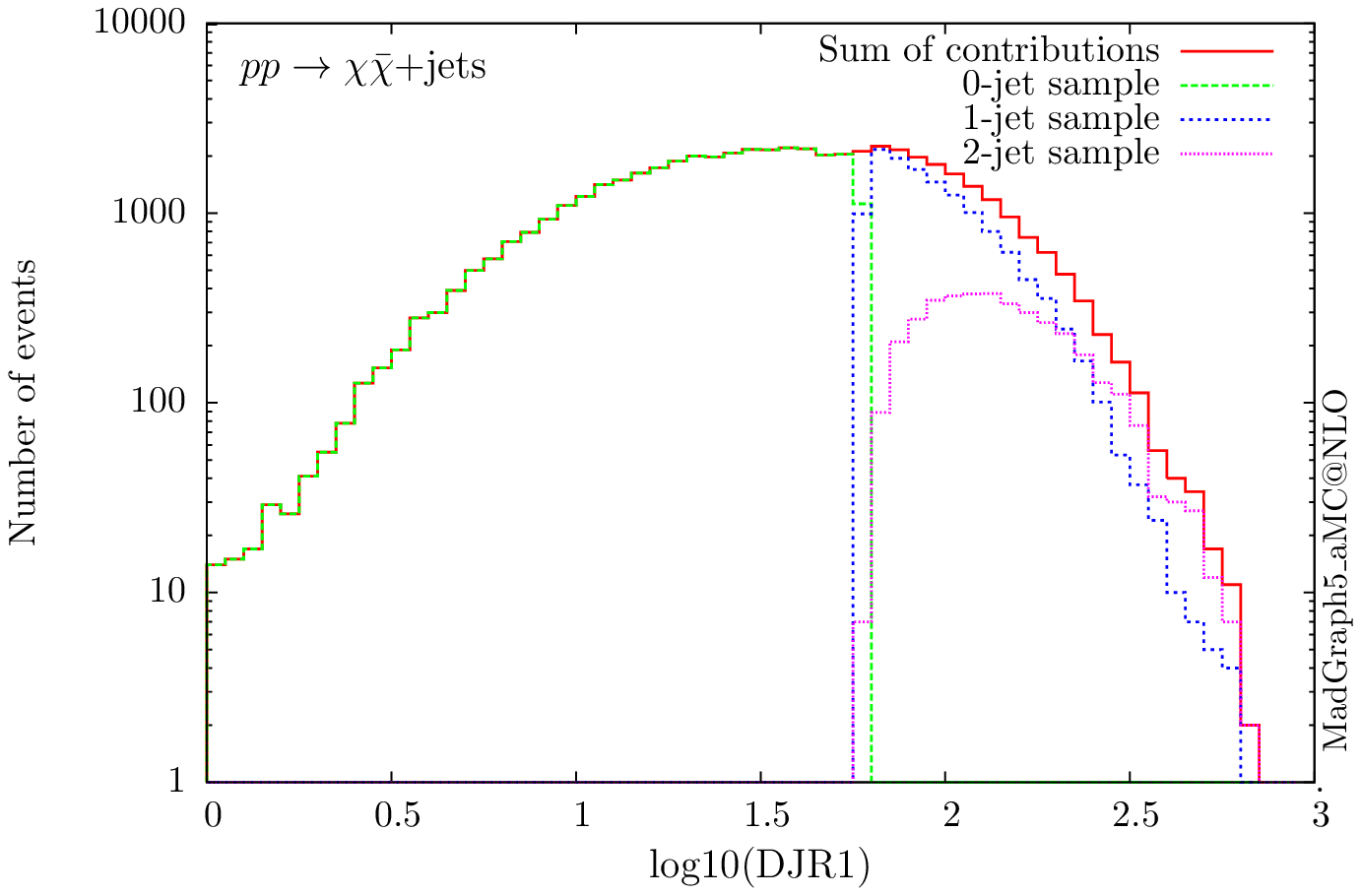}
\label{h1h1_H1}
\end{minipage}
\hspace{0.5cm}
 \begin{minipage}[t]{0.5\linewidth}
 \centering
 \includegraphics[trim=1.2cm 0 0 0,scale=0.56]{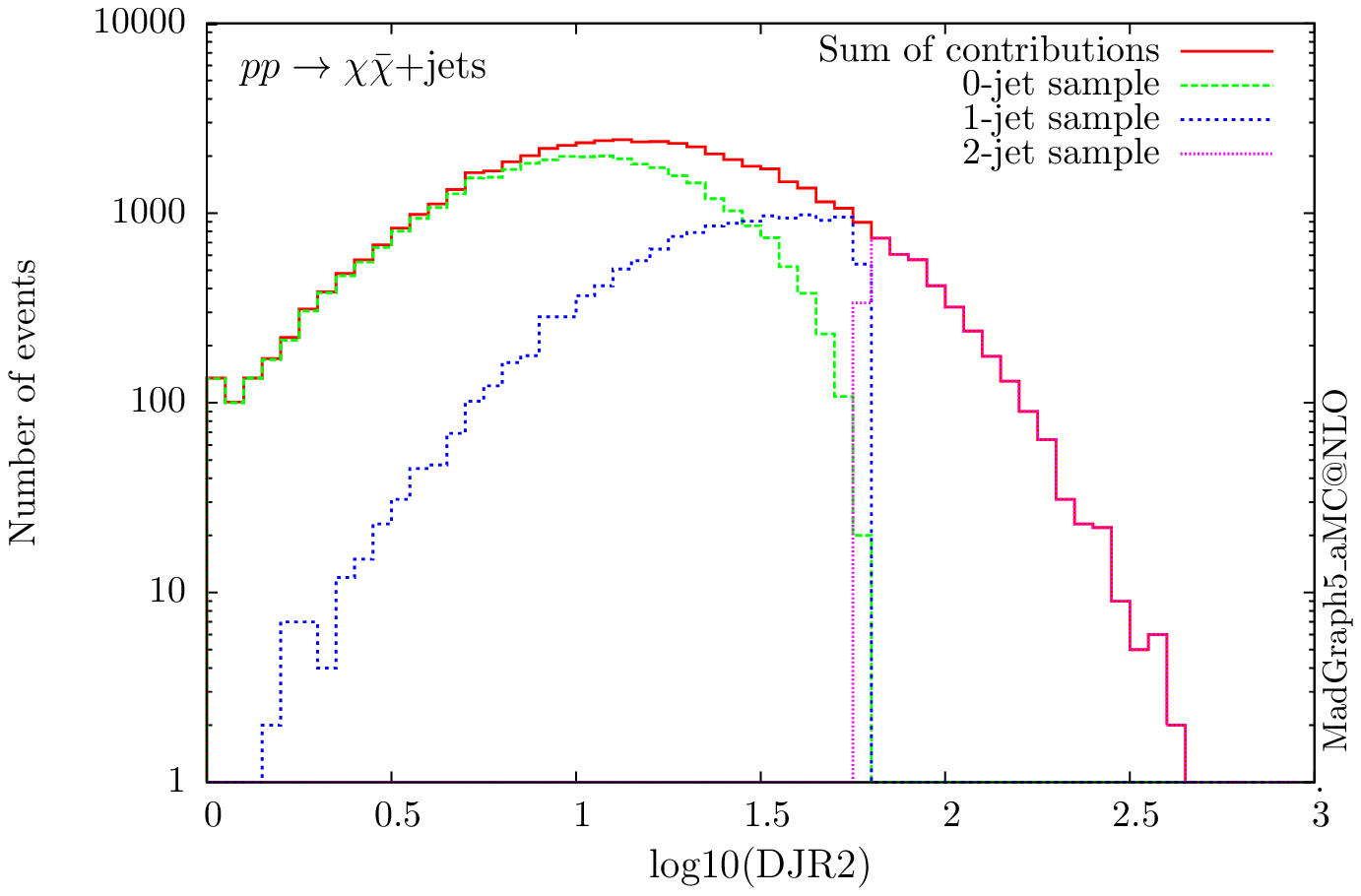}
 \end{minipage}
\caption{\label{fig:DJR} Differential jet rate distributions for a scalar mediator in the resonant DM production scenario, for a merging scale of 60 GeV.  } 
\end{figure}

We will compare results for the three scenarios presented above, showing the distributions for the most relevant kinematical variables. As an example of the checks performed to ensure the merging scale chosen was an appropriate one, we show for one process --the resonant scalar production-- the differential jet rate distributions in figure \ref{fig:DJR}. This distributions have been obtained with {\sc Pythia8} with a matching scale ($Q_{\textrm{cut}}$) of 60 GeV.

\begin{figure}[tb!]
\centering
\includegraphics[trim=1.4cm 0 0 0,scale=0.56]{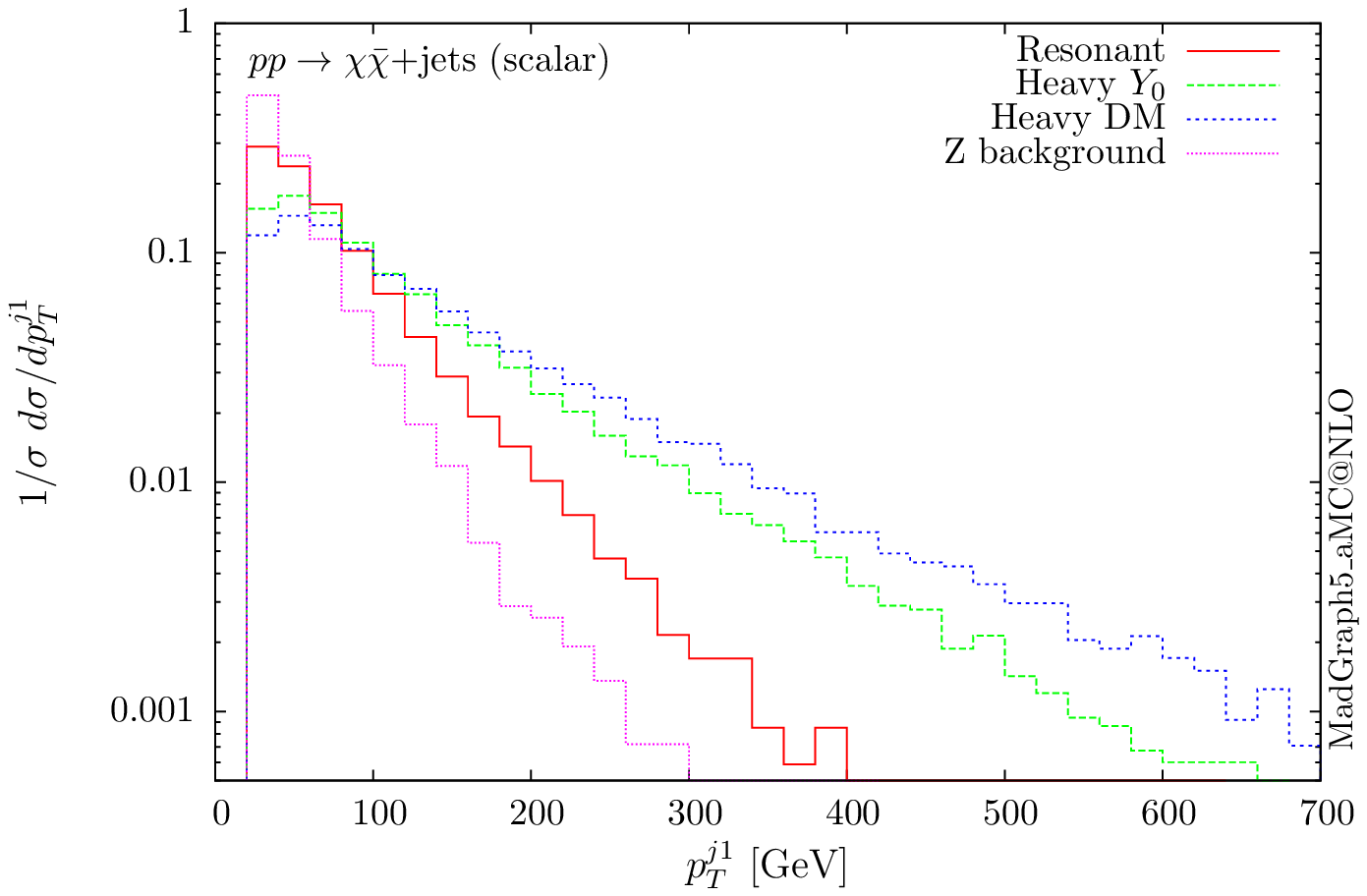}
\hspace{0.5cm}
 \centering
 \includegraphics[trim=1.2cm 0 0 0,scale=0.56]{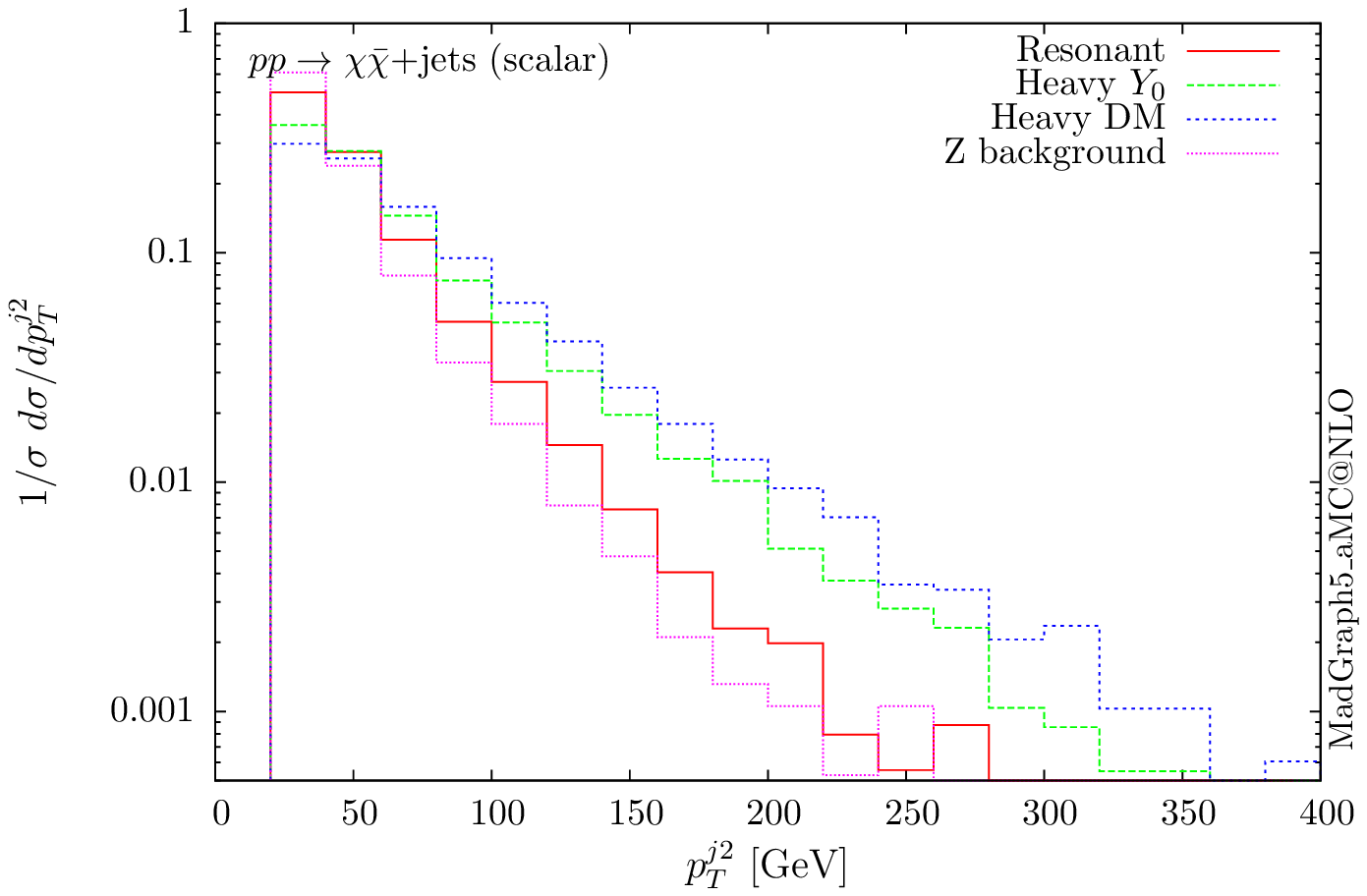}
\caption{\label{fig:jetspt} Hardest and second hardest jet transverse momentum distribution for $ p  p \to \chi \bar{\chi} +$ jets for a scalar mediator. } 
\end{figure}

\begin{figure}[h!]
\centering
\includegraphics[trim=1.2cm 0 0 0,scale=0.56]{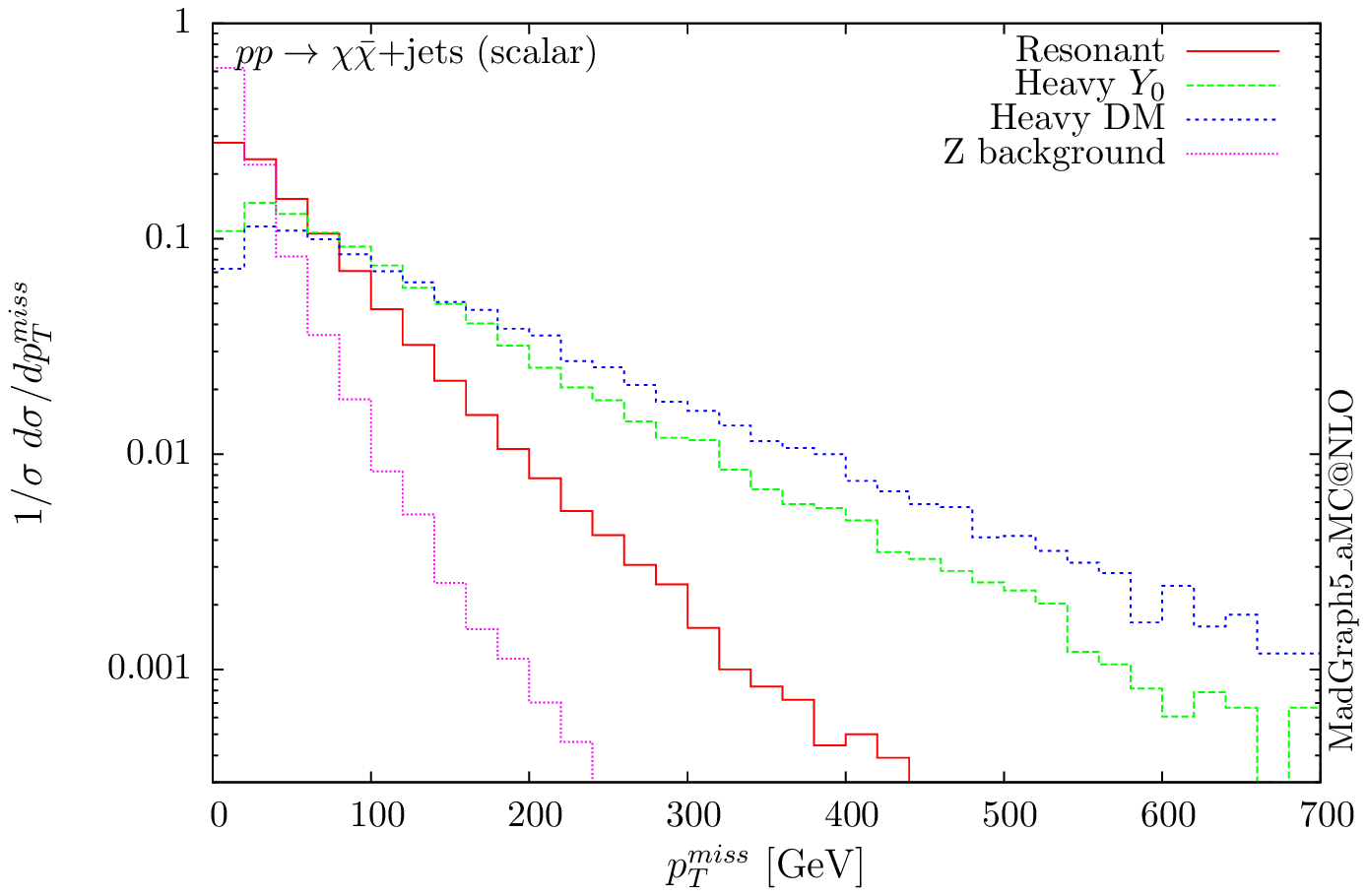}
\caption{Missing transverse momentum distribution for $ p  p \to \chi \bar{\chi} +$ jets for a scalar mediator.}
\label{misspt}
\hspace{0.5cm}
\centering
\includegraphics[trim=1.2cm 0 0 0,scale=0.56]{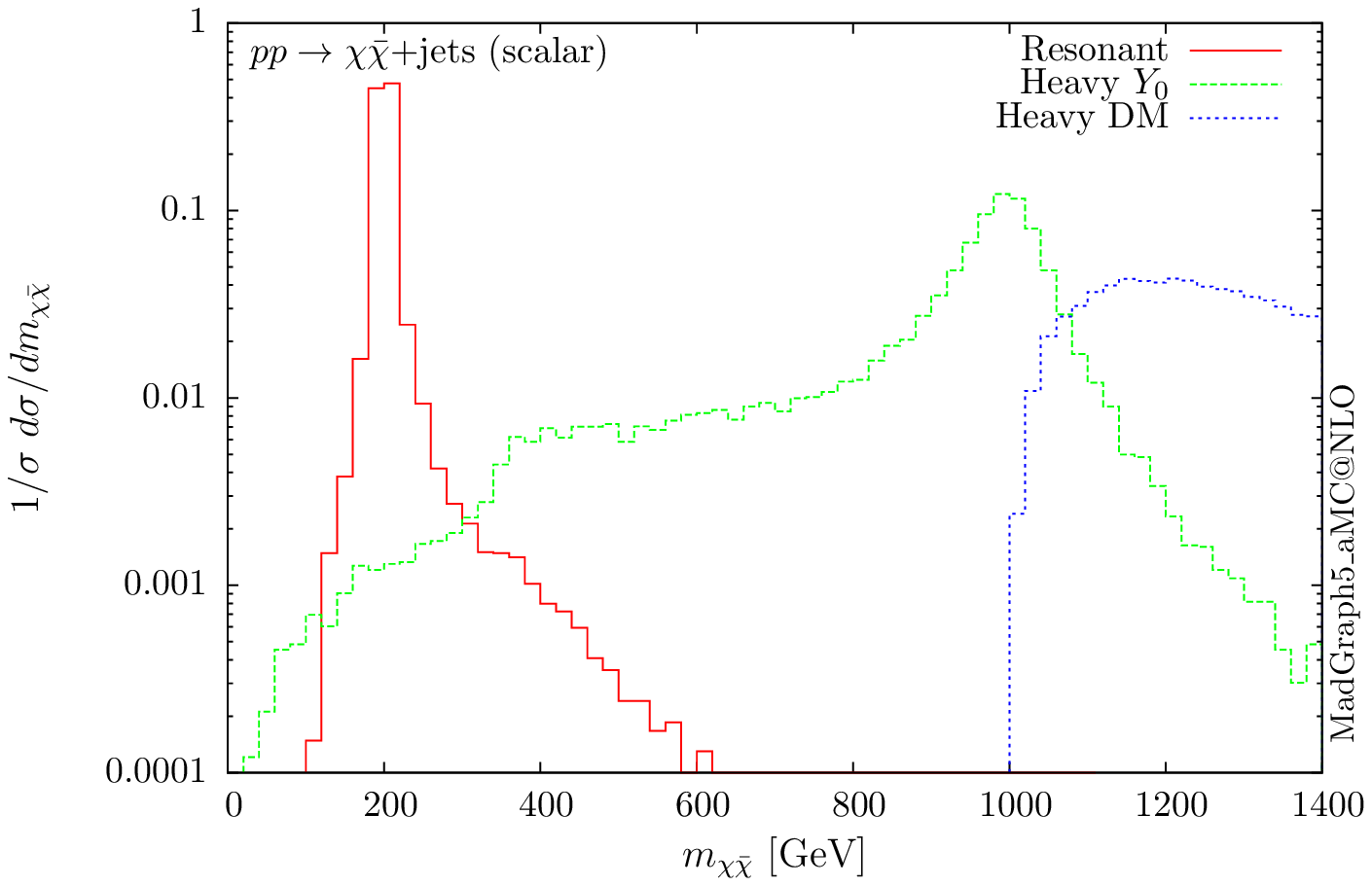}
\caption{Invariant mass distribution for the DM pair $ p  p \to \chi \bar{\chi} +$ jets for a scalar mediator.  }
\label{mass}
\end{figure}

The normalised distributions for the hardest, second hardest jet and the missing transverse momentum for the three scenarios are shown in Figs. \ref{fig:jetspt} and \ref{misspt} respectively.  The invariant mass distributions of the DM pair for the various benchmarks are presented in Fig. ~\ref{mass}.
The resonant curve displays a sharp resonant peak, while the heavy-mediator scenario has an important tail at low invariant masses, with the off-shell region contributing significantly to the cross-section. In both curves a threshold effect can be observed at $2m_t$ when the top quarks running in the loop become on-shell. We note that the mass of 1~TeV for the mediator is not sufficiently high for the EFT approach to be valid as the mass of the mediator is probed, as shown clearly in Fig. ~\ref{mass}. For the third and  final scenario --heavy DM-- the production threshold lies above the mediator mass and therefore no resonant structure arises.

We see that the $Z$+ jets background falls faster than any of the DM scenarios, for all the transverse momentum distributions shown. This implies that while inclusively it is overwhelming, deviations from the background can be observed more easily  by searching in the boosted regions. Comparing the three scenarios, we notice that the distributions for the resonant scenario, fall more rapidly, while the heavy DM and heavy mediator scenarios lead to harder distributions in the tails. 

\begin{figure}[tb!]
 \begin{minipage}[t]{0.5\linewidth}
\centering
\includegraphics[trim=1.2cm 0 0 0,scale=0.56]{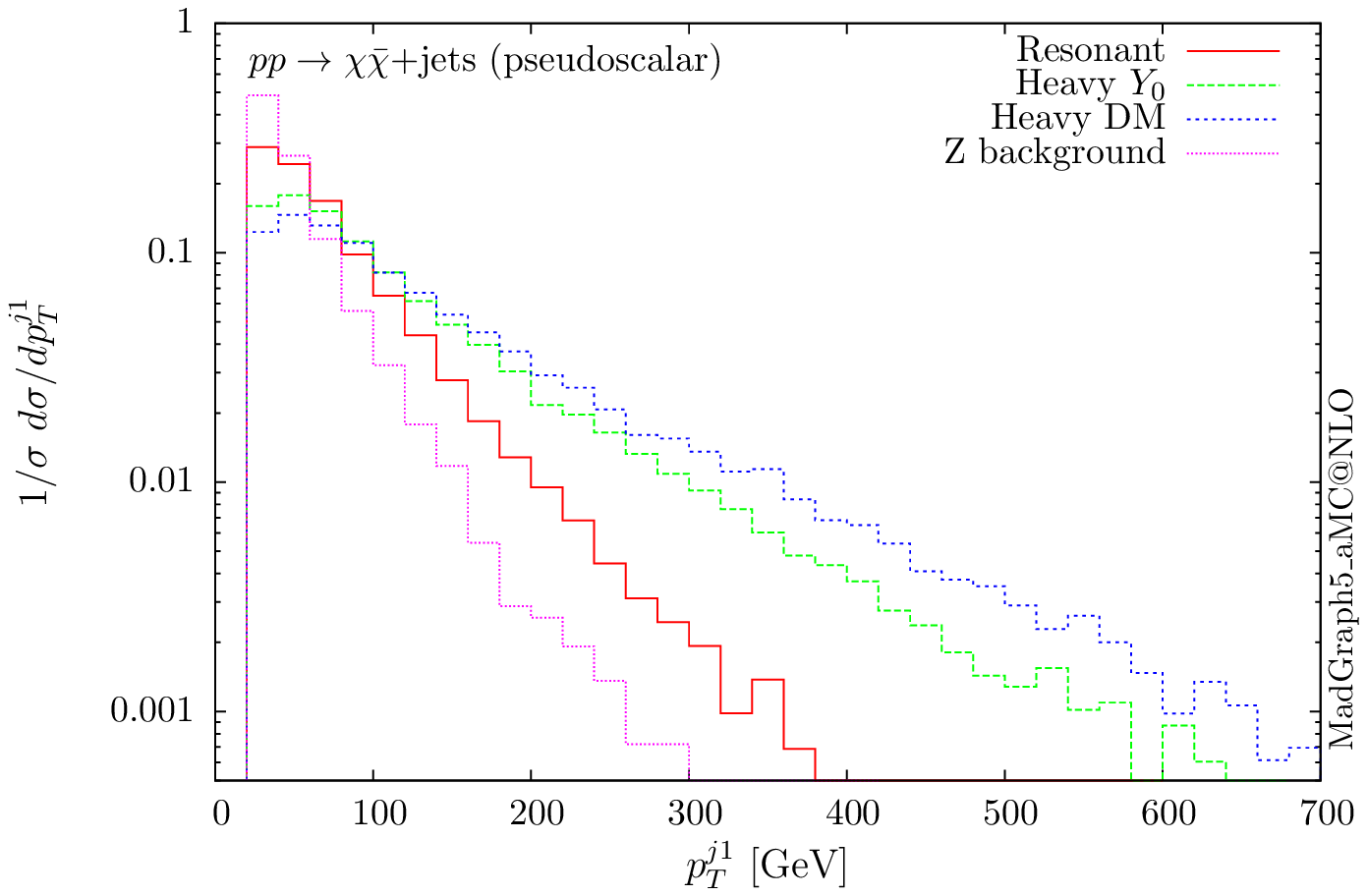}
\label{h1h1_H1}
\end{minipage}
\hspace{0.5cm}
 \begin{minipage}[t]{0.5\linewidth}
 \centering
 \includegraphics[trim=1.2cm 0 0 0,scale=0.56]{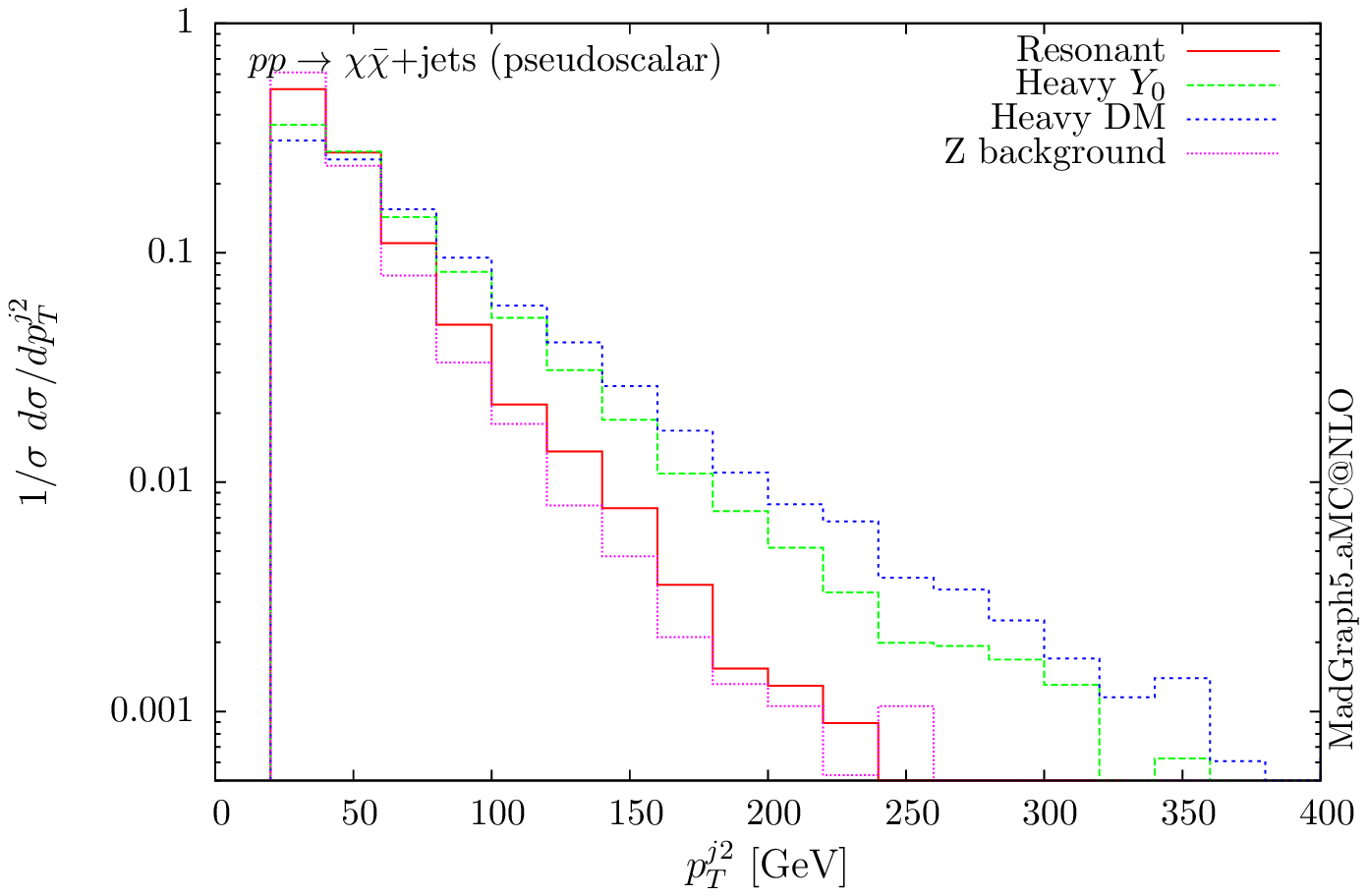}
 \end{minipage}
\caption{\label{fig:jetptP} Hardest and second hardest jet transverse momentum distribution for $ p  p \to \chi \bar{\chi} +$ jets for a pseudoscalar mediator. } 
\end{figure}

\begin{figure}[h!]
\centering
\includegraphics[trim=1.2cm 0 0 0,scale=0.56]{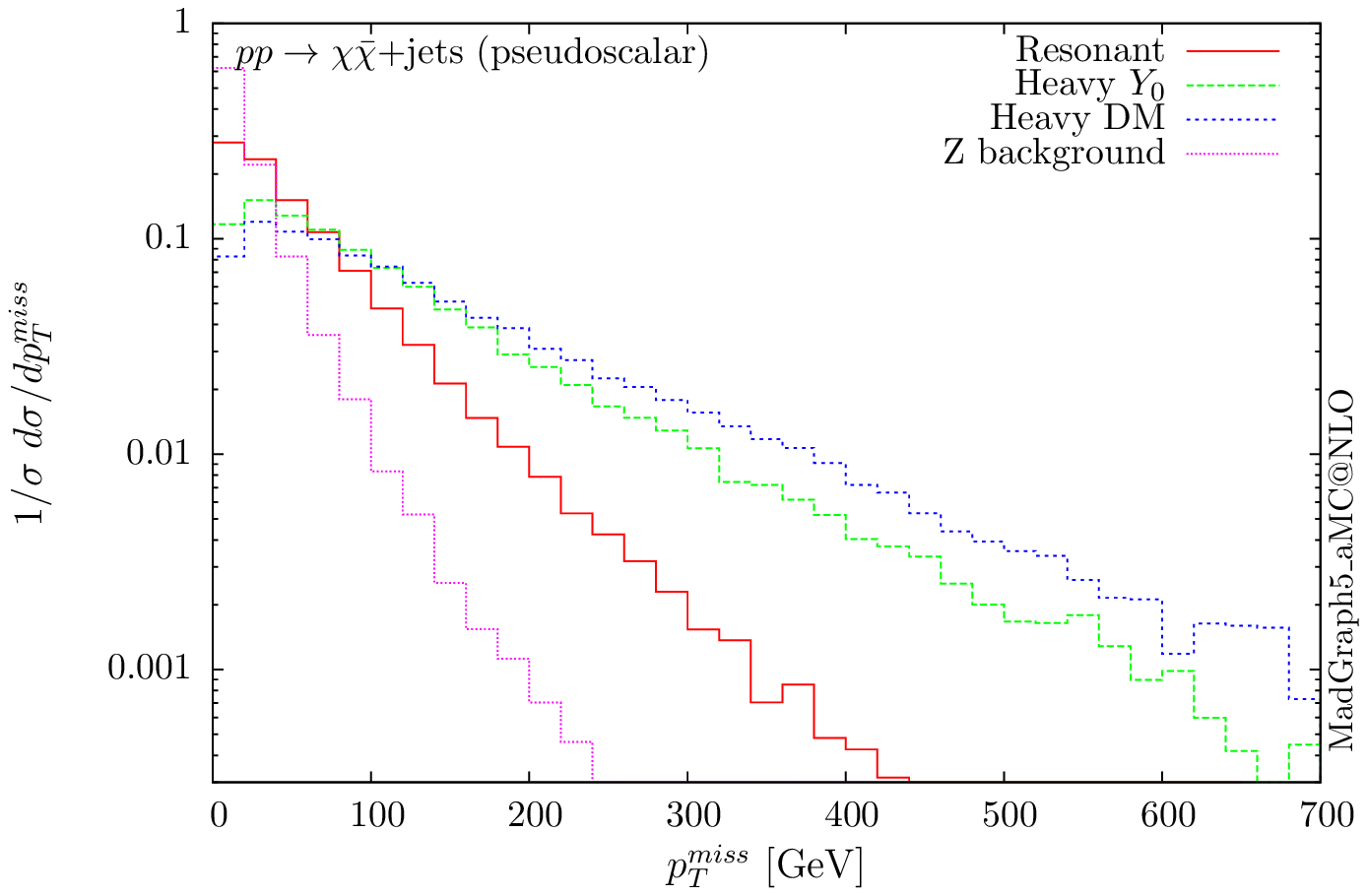}
\caption{Missing transverse momentum distribution for $ p  p \to \chi \bar{\chi} +$ jets for a pseudoscalar mediator. }
\label{missptP}
\hspace{0.5cm}
\centering
\includegraphics[trim=1.2cm 0 0 0,scale=0.56]{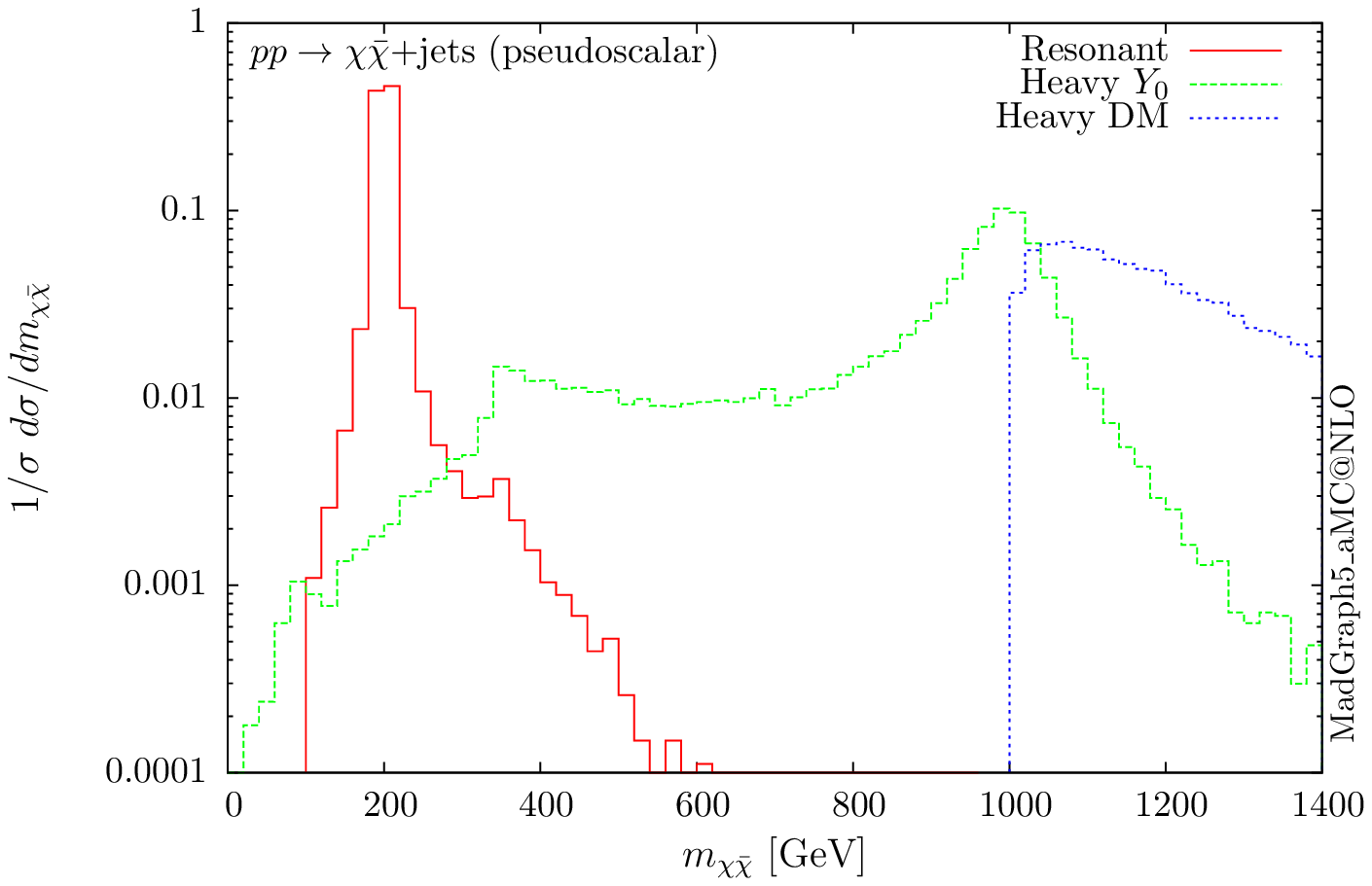}
\caption{Invariant mass distribution for the DM pair $ p  p \to \chi \bar{\chi} +$ jets for a pseudoscalar mediator.  }
\label{massP}
\end{figure}

The corresponding results for the pseudoscalar mediator are shown in Figs.~\ref{fig:jetptP}, \ref{missptP} and \ref{massP}. These are almost identical to the results of the scalar propagator, with no visible difference in the normalised distributions for the jet and missing transverse energy. We recall here that a variable that can be used to distinguish between scalar and pseudoscalar is the azimuthal separation between the two leading jets produced in this process, as discussed for the SM Higgs in $H+2$ jets in \cite{Demartin:2014fia} and for DM searches in \cite{Haisch:2013fla}. The slightly sharper threshold visible at $2m_t$ is due to the different behaviour of the scalar and pseudoscalar amplitudes. 

\begin{figure}[h!]
\centering
\includegraphics[trim=1.2cm 0 0 0,scale=0.56]{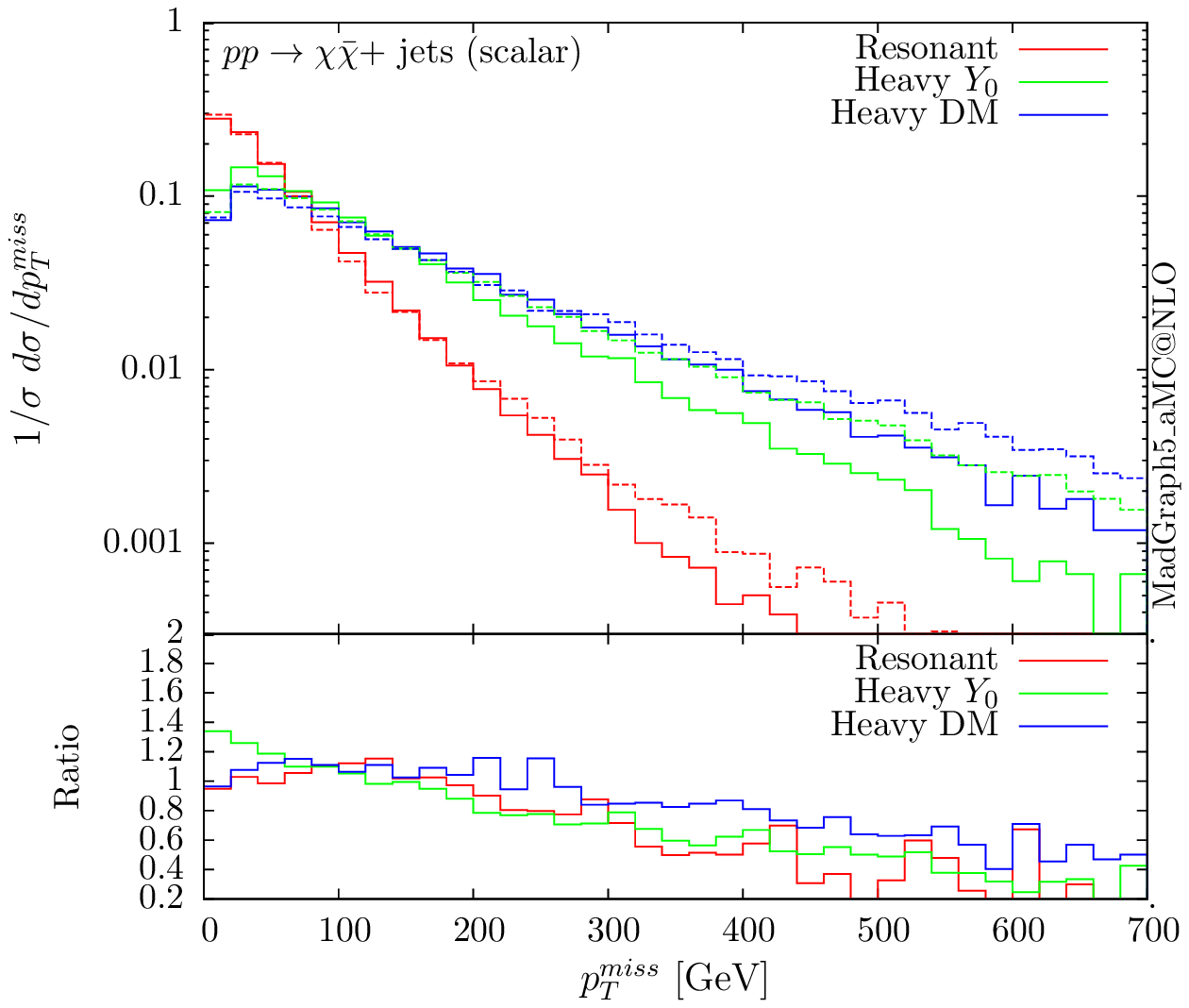}
\caption{Missing transverse momentum distribution for $ p  p \to \chi \bar{\chi} +$ jets for a scalar mediator, using the exact loops (solid lines) and the top EFT (dashed lines) and the corresponding ratio.  }
\label{topcomp}
\end{figure}

\vskip 0.5cm
For the case of scalar or pseudoscalar mediators, one could also consider production cross-sections in the infinite top mass limit, i.e. employing the Lagrangian of the form \cite{Dawson:1998py}: 
\begin{equation}
\mathcal{L}=\frac{\alpha_s}{12\pi v} g^S_{t} G_{\mu\nu} G^{\mu\nu} Y_0+\frac{\alpha_s}{8\pi v} g^P_{t} G_{\mu\nu} \tilde{G}^{\mu\nu} Y_0.
\end{equation} 
Such an approach has the clear advantage of being much simpler than computing full loop amplitudes and thanks to this simplicity, the possibility of including NLO corrections in QCD, see for instance \cite{Demartin:2014fia}. Nevertheless, the accuracy of the infinite top approximation with respect to the exact loop computation needs to be assessed on a case-by-case basis. 
As an example, we show, in Fig. \ref{topcomp},  a comparison for the missing transverse energy distribution in the case of the scalar mediator.
 It is clear from the plot that integrating out the top quark leads to harder distributions in the tails and  the top-EFT result overshoots the loop-one for all three scenarios. We see that for the resonant case, the infinite top mass limit provides a reliable prediction of the distribution shape up to 200 GeV. These observations are qualitatively  consistent with the corresponding studies for Higgs production in the SM, where the infinite top mass limit fails at high Higgs transverse momentum. Considering the fact that the DM searches focus on the boosted regions to overcome the large SM backgrounds, to avoid overestimating the signal and consequently setting inaccurate limits on the various DM model parameters, one needs to resort to the loop computation. 

\subsection{Differential results for the vector and axial-vector mediators}
\begin{figure}[tb!]
 \begin{minipage}[t]{0.5\linewidth}
\centering
\includegraphics[trim=1.2cm 0 0 0,scale=0.56]{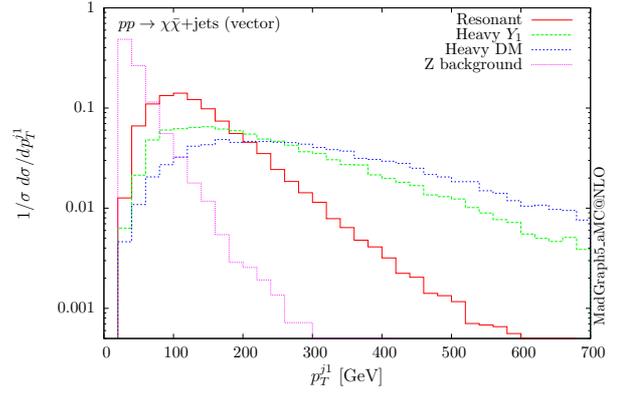}
\label{h1h1_H1}
\end{minipage}
\hspace{0.5cm}
 \begin{minipage}[t]{0.5\linewidth}
 \centering
 \includegraphics[trim=1.2cm 0 0 0,scale=0.56]{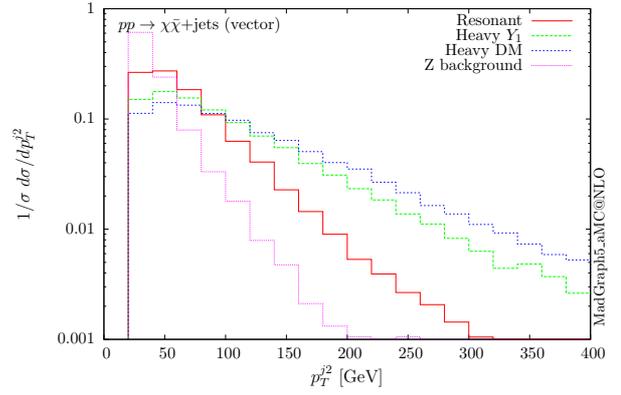}
 \end{minipage}
\caption{\label{fig:jetptV} Hardest and second hardest jet transverse momentum distribution for $ p  p \to \chi \bar{\chi} +$ jets for a vector mediator. } 
\end{figure}

\begin{figure}[h!]
\centering
\includegraphics[trim=1.2cm 0 0 0,scale=0.56]{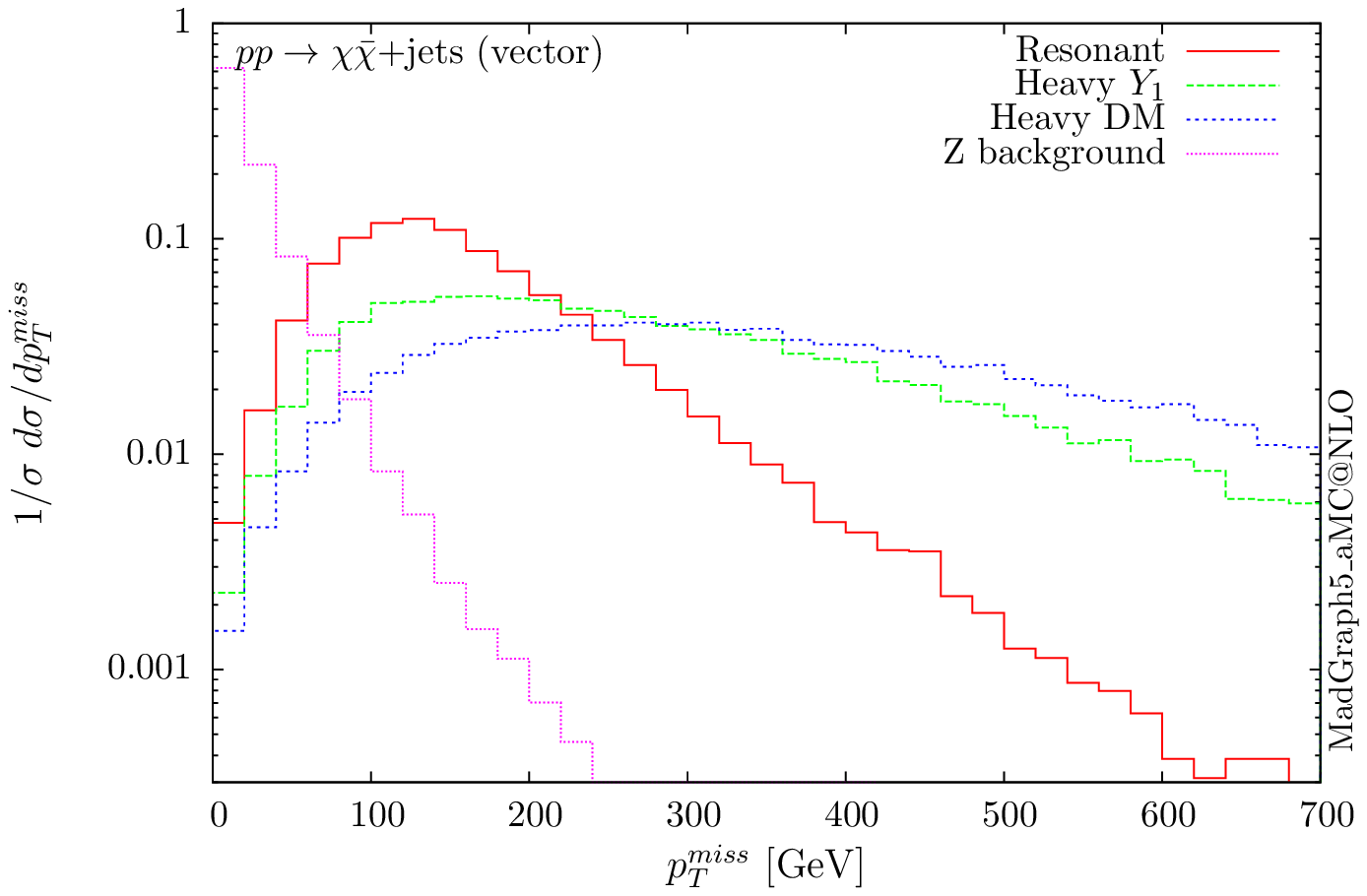}
\caption{Missing transverse momentum distribution for $ p  p \to \chi \bar{\chi} +$ jets for a vector mediator. }
\label{misspt_V}
\hspace{0.5cm}
\centering
\includegraphics[trim=1.2cm 0 0 0,scale=0.56]{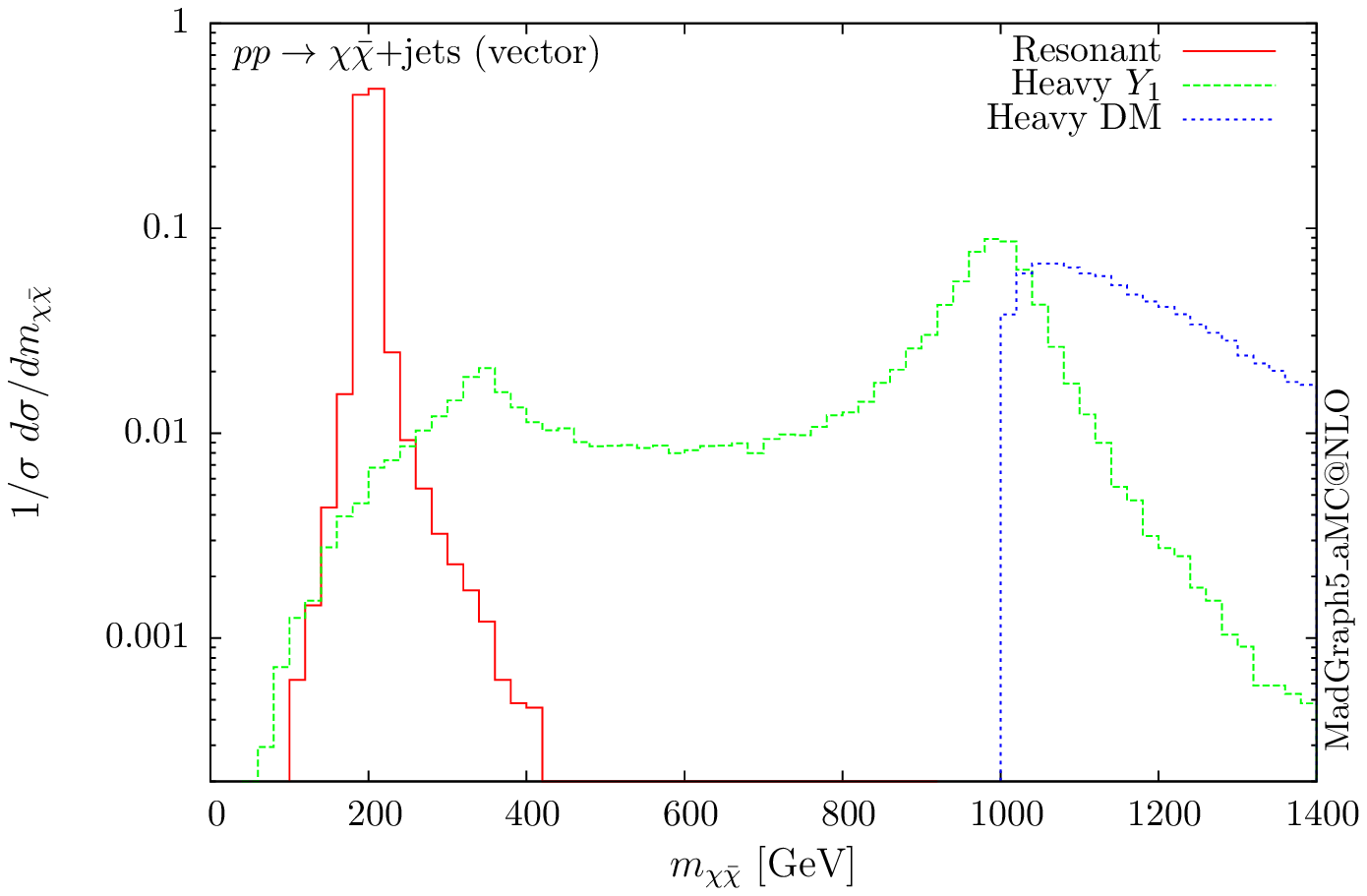}
\caption{Invariant mass distribution for the DM pair $ p  p \to \chi \bar{\chi} +$ jets for a vector mediator.   }
\label{massV}
\end{figure}

Following the same procedure, one can obtain results for the vector and axial vector mediators. When merging samples of different multiplicities, one needs to take into account that for a vector mediator, the 0-jet contribution is zero due to  Landau-Yang theorem~\cite{Landau:1948kw,Yang:1950rg}. The quark-gluon initiated contributions to the 1-jet process also vanish. In fact, any diagram involving a triangle with two gluons and a vector mediator as the external legs vanishes, due to charge conjugation invariance. Therefore, in the vector case, we only merge the 1 and 2-jet contributions. For numerical stability we have removed the triangle diagrams at the time of generation of the code. The differential distributions are shown in Figs. \ref{fig:jetptV}, \ref{misspt_V} and \ref{massV}. 

The three scenarios display the same pattern as for the scalar and pseudoscalar, i.e. the resonant case giving the most rapidly falling distributions. The difference in the shape between the vector and scalar/pseudoscalar distributions originates from the absence of the 0-jet sample. Consequently,  both the missing $p_T$ and hardest jet distributions go to zero at low $p_T$ and present a maximum at  rather large $p_T$ values. The distribution of the invariant mass of the DM pair for the heavy mediator scenario shows a peak at both the $2m_t$ threshold and the mass of the mediator.

Similar results can be obtained for the axial-vector mediator. In this case though, there is a subtlety related to the fact that if the axial-vector mediator couples only to the top quark, this will lead to a gauge anomaly. In the SM the gauge anomaly for the $Z$ is exactly cancelled, as the axial vector coupling to the $Z$ for up and down quarks differs by a minus sign. Only coupling the mediator to the top, would imply that the theory is anomalous, which is not expected in any UV complete theory. As we have already discussed in the introduction, a minimal solution to this problem is to allow in this particular case the axial-vector mediator to couple also to the bottom quark, with a coupling opposite in sign to that of the top.  In this case, we perform the computation in the 4F scheme to again focus on the loop-induced production.    

\begin{figure}[tb!]
 \begin{minipage}[t]{0.5\linewidth}
\centering
\includegraphics[trim=1.2cm 0 0 0,scale=0.56]{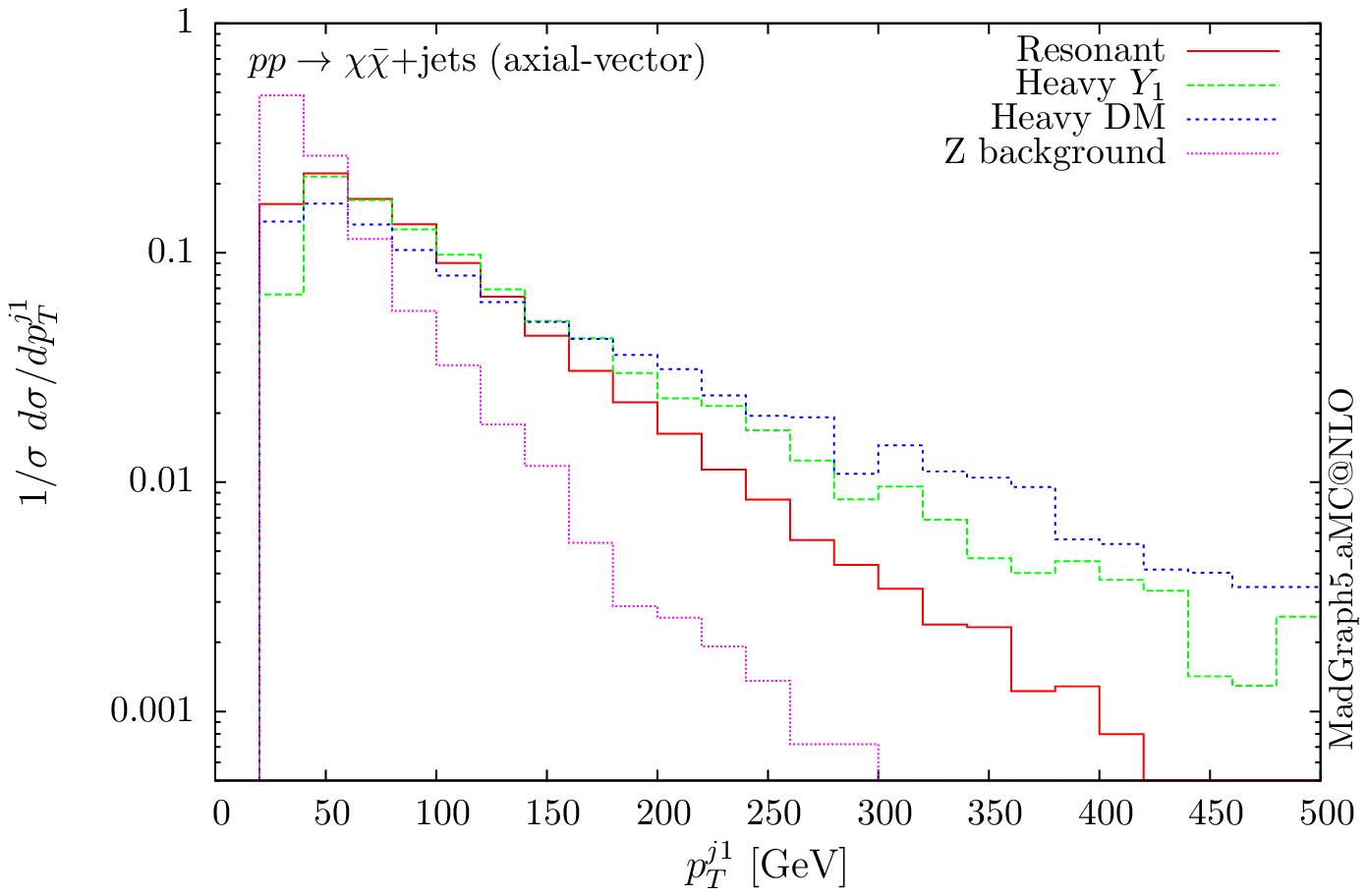}
\label{h1h1_H1}
\end{minipage}
\hspace{0.5cm}
 \begin{minipage}[t]{0.5\linewidth}
 \centering
 \includegraphics[trim=1.2cm 0 0 0,scale=0.56]{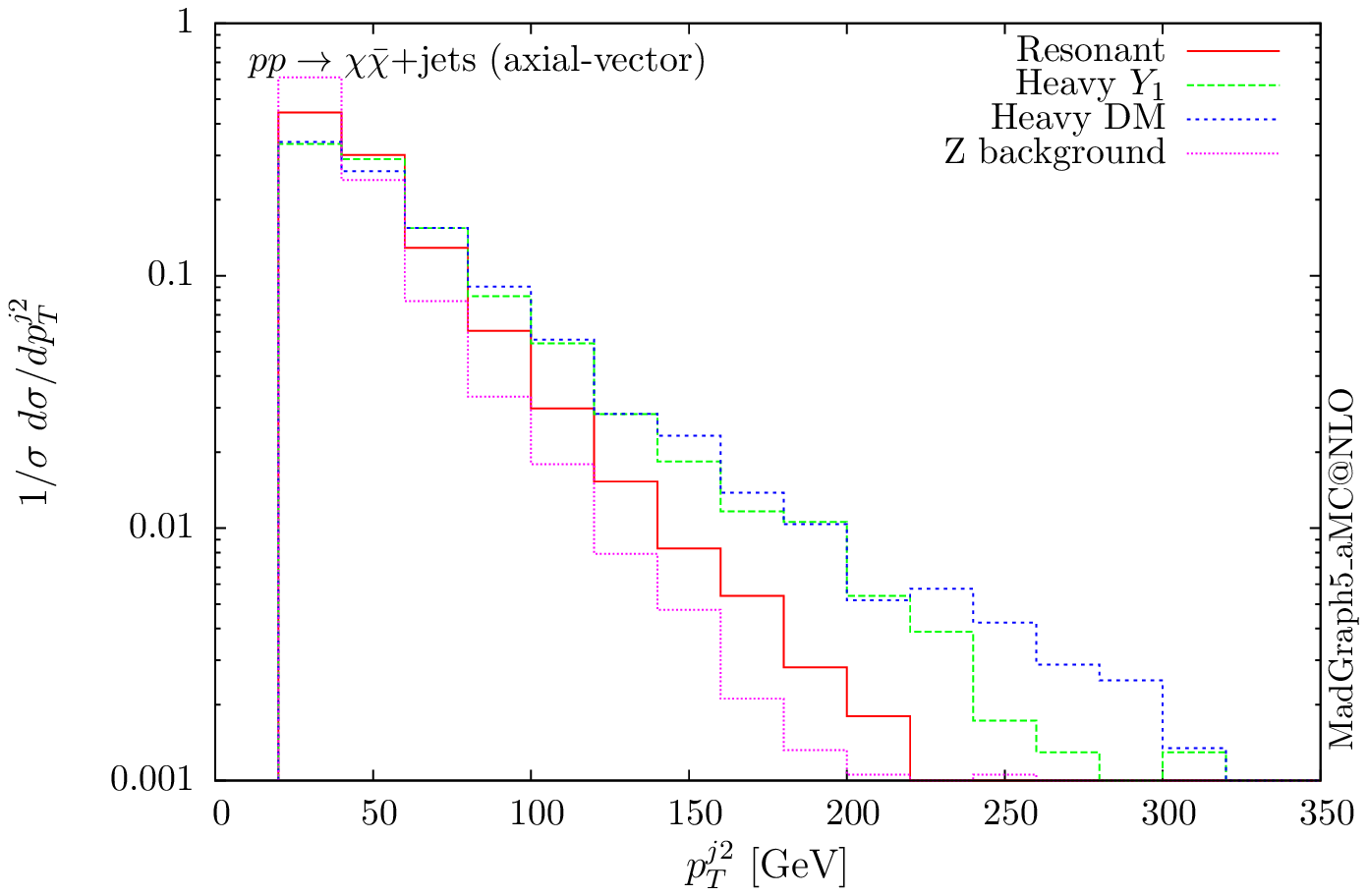}
 \end{minipage}
\caption{\label{fig:jetpta} Hardest and second hardest jet transverse momentum distribution for $ p  p \to \chi \bar{\chi} +$ jets for an axial-vector mediator. } 
\end{figure}

\begin{figure}[h!]
\centering
\includegraphics[trim=1.2cm 0 0 0,scale=0.56]{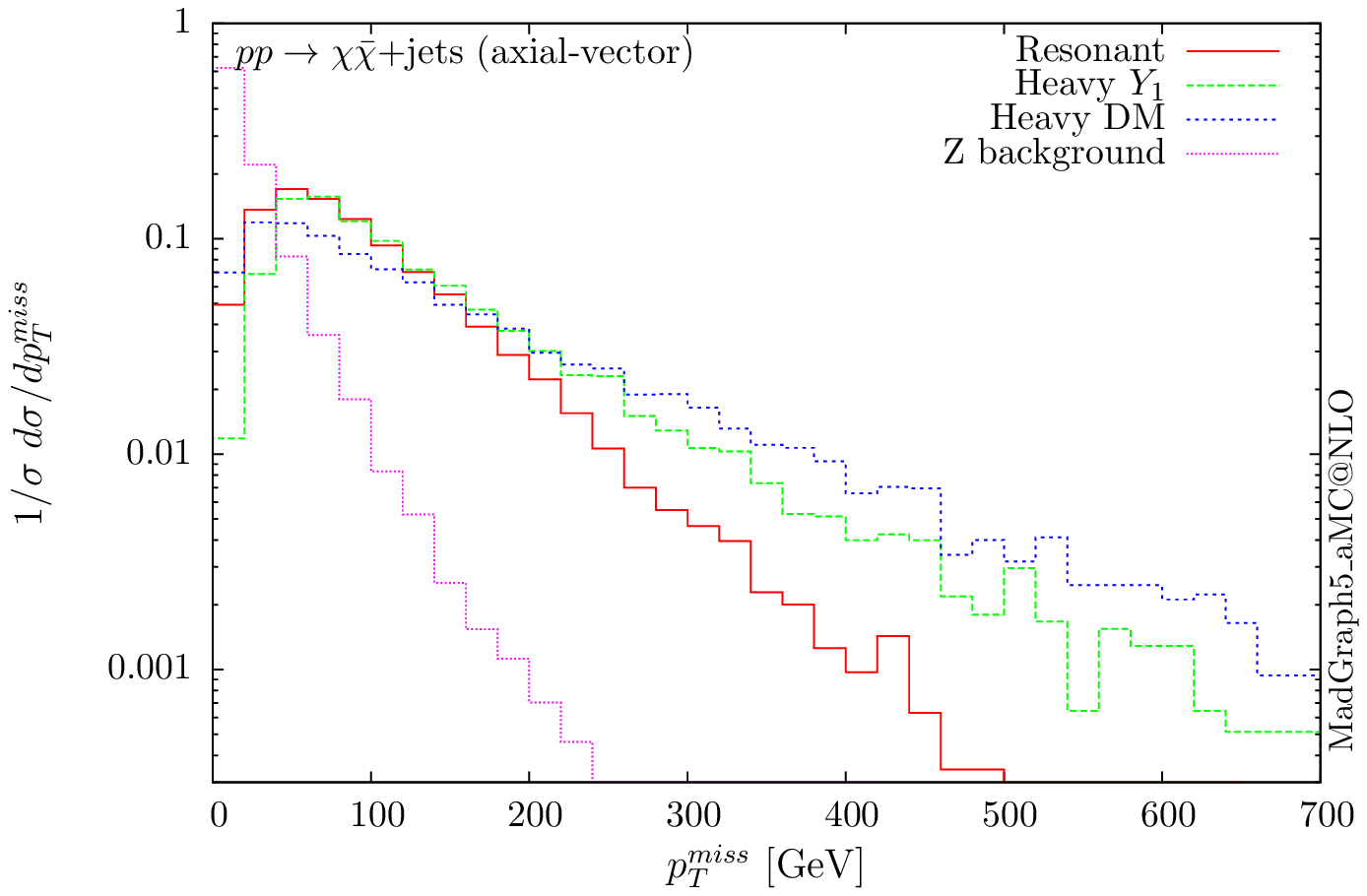}
\caption{Missing transverse momentum distribution for $ p  p \to \chi \bar{\chi} +$ jets for an axial-vector mediator. }
\label{misspt_a}
\hspace{0.5cm}
\centering
\includegraphics[trim=1.2cm 0 0 0,scale=0.56]{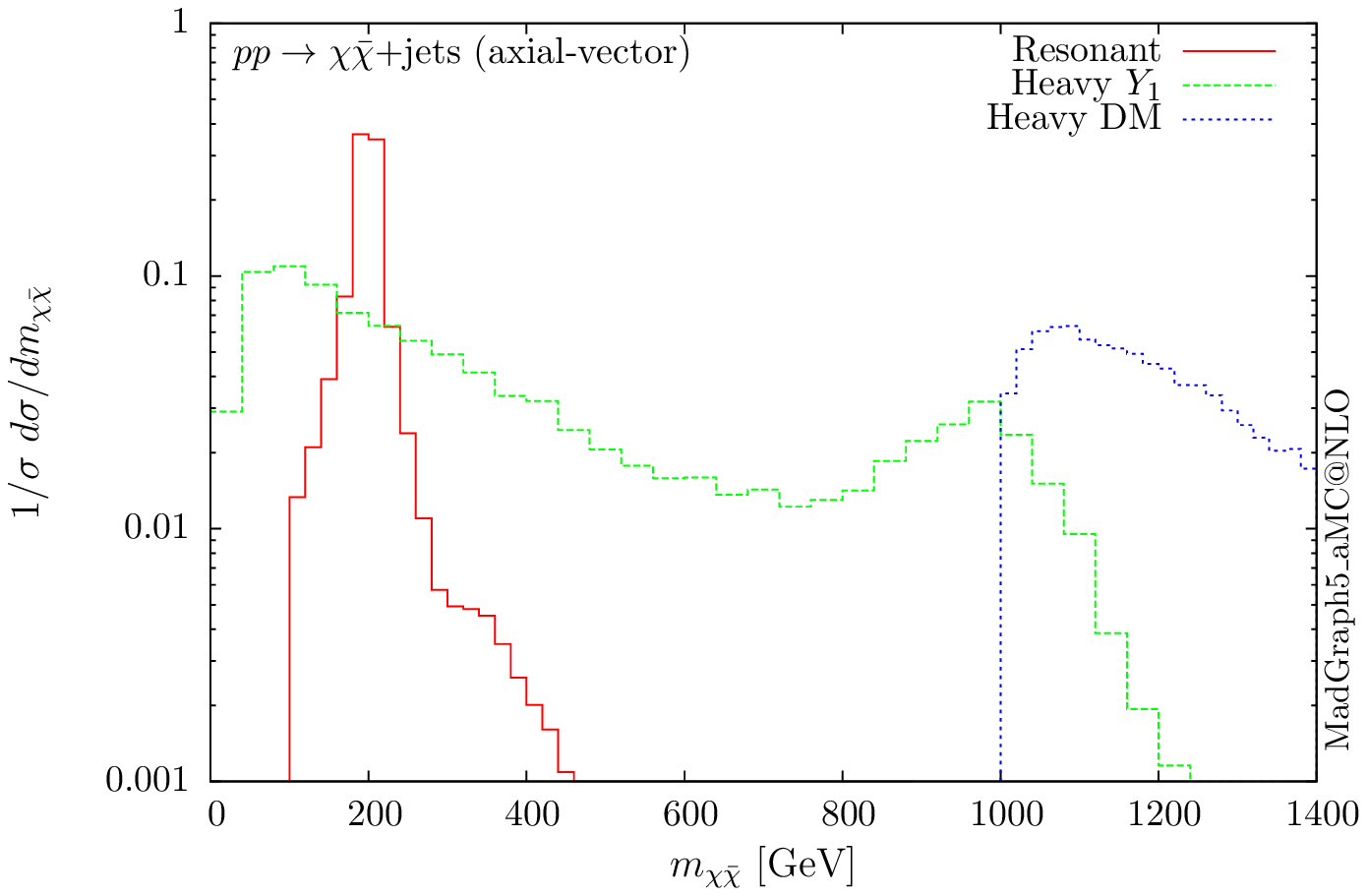}
\caption{Invariant mass distribution for the DM pair $ p  p \to \chi \bar{\chi} +$ jets for an axial-vector mediator.   }
\label{massA}
\end{figure}

The differential distributions are shown in Figs. \ref{fig:jetpta}, \ref{misspt_a} and \ref{massA}. The relative shapes for the three scenarios follow the same patterns as the other three couplings for the transverse momentum distributions, i.e. the resonant one falls more rapidly. A difference compared to the vector mediator is seen in the DM invariant mass distribution for the heavy mediator scenario. Due to the presence of the bottom quarks in the loops, a significant fraction of the cross-section lies at low invariant masses.  The fact that the width of the axial-vector mediator is significantly larger, further enhances the off-shell contribution. In fact, both the transverse momentum and missing energy distributions are softer than the corresponding ones for the vector mediator. Nevertheless, these distributions remain harder than those of the $Z$ background, facilitating boosted technique based searches.

\section{Other mono-X signals}
\label{monoX}
\begin{figure}[h!]
\centering
\includegraphics[trim=1cm 0 0 0,scale=0.45]{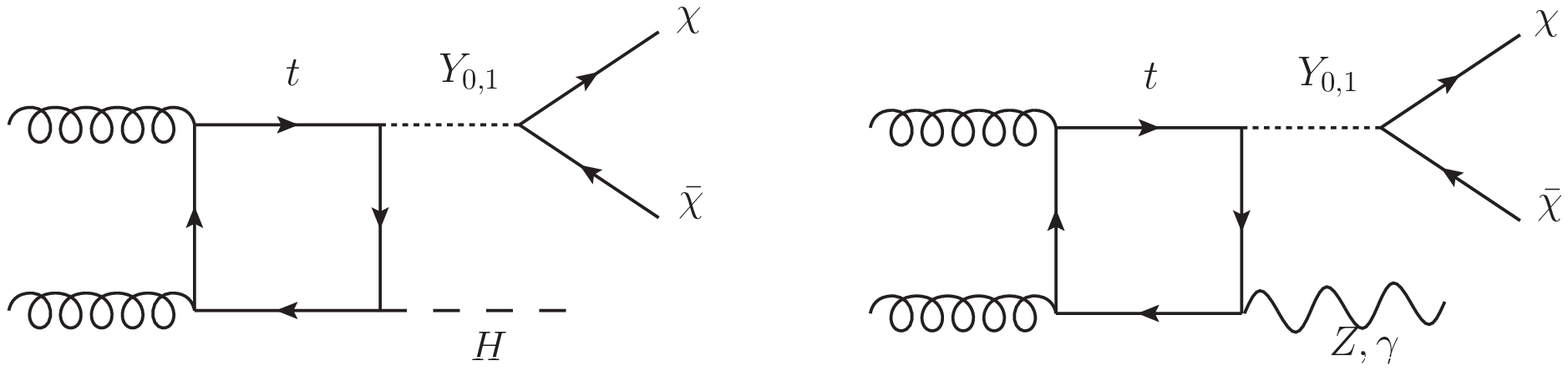}
\caption{Representative Feynman diagrams contributing to the mono-X processes in the simplified model.}
\label{fig:monoX}
\end{figure}

In this section, we consider the possibility of other mono-X signals within our simplified model. Three cases are considered: mono-$Z$, mono-Higgs and mono-photon. In all cases the production mode is through top quark loops in gluon fusion. The only contribution is via box diagrams as the mediators only couple to the top quark.  Sample diagrams are shown in Fig.~\ref{fig:monoX}. We study the same three benchmark points discussed in the previous section. However, in this section, we limit ourself for simplicity to the parton level, and focus on some general physics considerations. 

\subsection{mono-Z}
\begin{table*}[t]
\renewcommand{\arraystretch}{1.3}
\begin{center} 
    \begin{tabular}{ l ? c | c | c | c  }
        \hline \hline
    Benchmark   & scalar         & pseudoscalar   & vector      & axial-vector\\        \hline
Resonant & $2.99\cdot 10^{-2} \, {}^{+36\%}_{-25\%}\,{}^{+1.3\%}_{-1.4\%}$ &  $3.28\cdot 10^{-2} \, {}^{+36\%}_{-25\%}\,{}^{+1.3\%}_{-1.4\%}$ & 3.26\cdot $10^{-3} \, {}^{+35\%}_{-24\%}\,{}^{+1.1\%}_{-1.3\%}$ & 8.98\cdot $10^{-2} \, {}^{+35\%}_{-24\%}\,{}^{+1.1\%}_{-1.2\%}$ \\ 
Heavy mediator & 2.20\cdot $10^{-4} \, {}^{+43\%}_{-28\%}\,{}^{+2.5\%}_{-2.5\%}$ & $2.08\cdot 10^{-4} \, {}^{+43\%}_{-28\%}\,{}^{+2.6\%}_{-2.5\%}$ & 2.15\cdot $10^{-6} \, {}^{+42\%}_{-28\%}\,{}^{+2.2\%}_{-2.3\%}$ & 1.52\cdot $10^{-4} \, {}^{+41\%}_{-27\%}\,{}^{+2.0\%}_{-2.0\%}$\\
Heavy DM & $4.75\cdot 10^{-7} \, {}^{+45\%}_{-29\%}\,{}^{+3.5\%}_{-3.4\%}$ & $1.40\cdot 10^{-6} \, {}^{+44\%}_{-28\%}\,{}^{+3.2\%}_{-3.1\%} $ & 1.05\cdot $10^{-8} \, {}^{+44\%}_{-28\%}\,{}^{+3.1\%}_{-3.0\%}$ &1.10\cdot $10^{-5} \, {}^{+44\%}_{-28\%}\,{}^{+3.3\%}_{-3.2\%}$ \\
\hline \hline
\end{tabular}
 \caption{\label{tab:monoZ} Cross sections (in pb) for gluon induced mono-Z production
   at the LHC at $\sqrt{s} = 13$~TeV for three mass benchmarks. A technical cut of 2~GeV has been set on the transverse momentum of all final heavy states. No Z branching ratios are included. }  
\end{center} 
\end{table*}
We start by considering the mono-$Z$ associated production. The total cross-sections for $p p \to \chi \bar{\chi} Z$  for unit couplings at 13 TeV are given in Table \ref{tab:monoZ}. We find that the scalar and pseudoscalar mediators give results of the same order of magnitude. Note that, for both scalar and pseudoscalar mediator production, it is only the axial-vector coupling of the top to the $Z$ boson which contributes, due to charge conjugation invariance. 

Similarly to the multi-jet case, the axial-vector mediator gives much larger cross-sections than the vector one. As discussed in Section \ref{subsec_xsec}, this is due to the propagator term $p^{\mu}p^{\nu}/M_{Y_1}^2$ which, when contracted with the axial-vector current, leads to terms proportional to $m_t m_{\chi}/M_{Y_1}^2$ potentially divergent in the limit of $M_{Y_1}\to 0$, due to the non-conservation of the axial current. 

\begin{figure}[h!]
\centering
\includegraphics[trim=1cm 0 0 0,clip,scale=0.56]{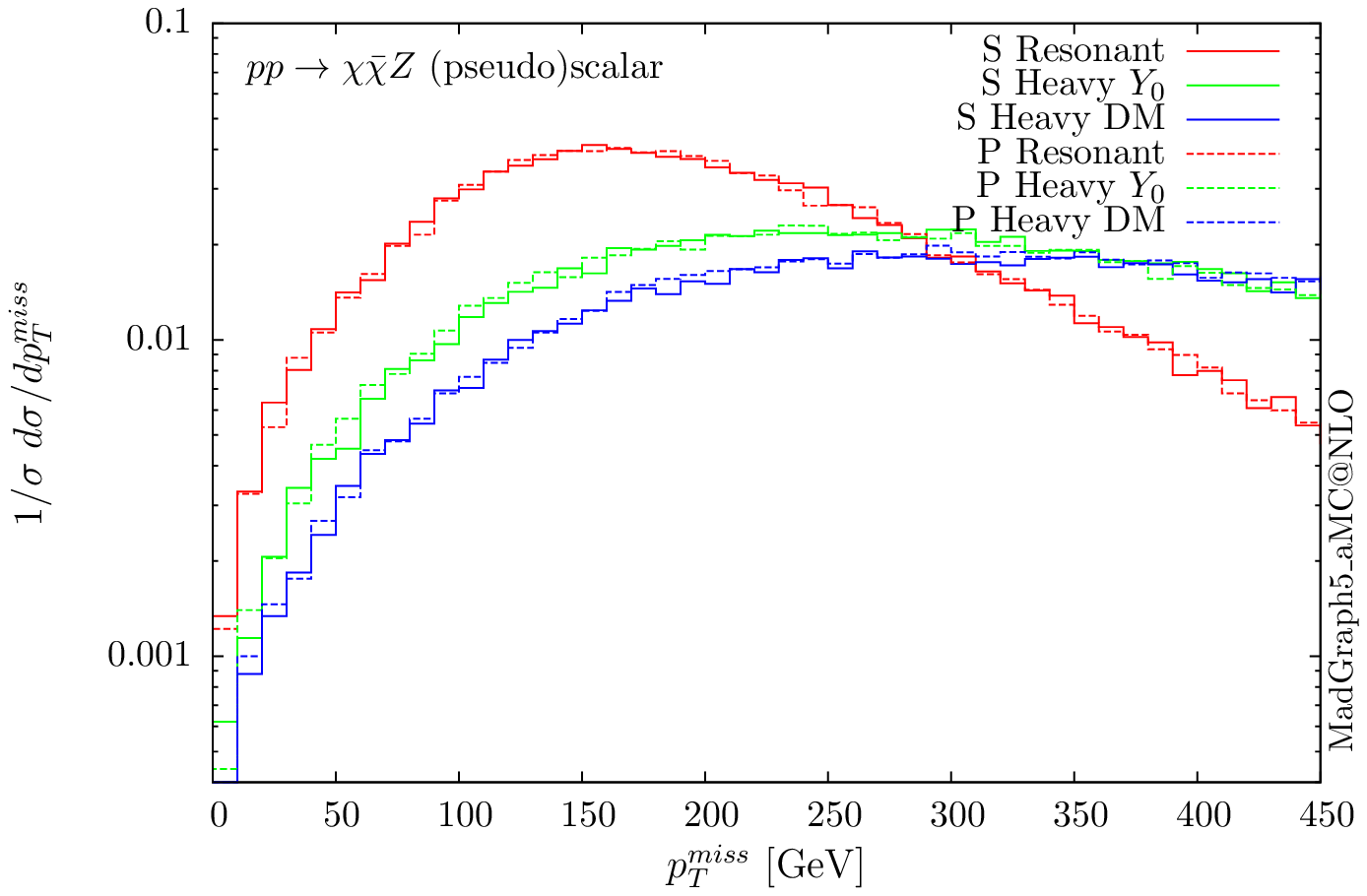}
\caption{Missing $p_T$ distribution for $ p  p \to \chi \bar{\chi} Z$ for the scalar (S) and pseudoscalar (P) mediators.}
\label{Zpt_scalar}
\end{figure}

\begin{figure}[h!]
\centering
\includegraphics[trim=1cm 0 0cm 0,scale=0.56]{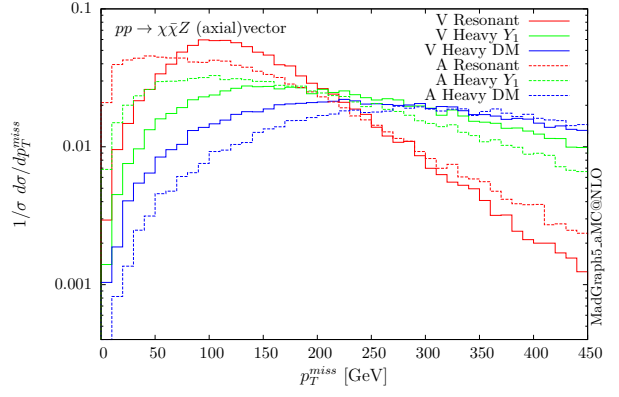}
\caption{Missing $p_T$ distribution for $ p  p \to \chi \bar{\chi} Z$ for the vector (V) and axial-vector (A) mediators. }
\label{Zpt_vector}
\end{figure}

The normalised parton-level missing transverse momentum distributions, i.e. the transverse momentum of the mediator, for the four types of mediator couplings are shown in Figs.~\ref{Zpt_scalar} and \ref{Zpt_vector}. Again the resonant scenario gives a distribution which falls more rapidly, and we find that the shape of the distributions are the same for the scalar and pseudoscalar mediators. These shapes are similar to those given by the box contribution to the $g g \to H Z$ production is the SM, as shown in \cite{Hespel:2015zea}. In the SM though, there is a significant, nearly exact at high $p_T$,  cancellation between the box and triangle diagrams, which leads to suppressed tails. This cancellation is absent in the DM scenarios, and the distributions fall very slowly. We note that the total cross-section  for the resonant case is smaller than the SM one, even though the masses ($m_H\sim 2 m_{\chi}$) are similar, as it is the triangle contribution that is the dominant one in the SM at low energies. 

Distributions corresponding to vector and axial-vector mediators  differ not only in the normalisation but also in the shapes for the resonant and heavy DM scenarios. From charge conjugation invariance we know that there is no mixed term $c_a c_v$  ($c_a/c_v$ are respectively the axial/vector couplings of the Z boson), and only the vector and axial-vector Z couplings contribute for the vector and axial mediators respectively. As discussed in the original computation of $g g  \to Z Z$ in \cite{Glover:1988rg} the contributions from $c_v^2$ and $c_a^2$ are different and lead to enhancements for massive quarks due to the non-conservation of the axial current.

\begin{figure}[h!]
\centering
\includegraphics[trim=1.2cm 0 0 0 cm,scale=0.56]{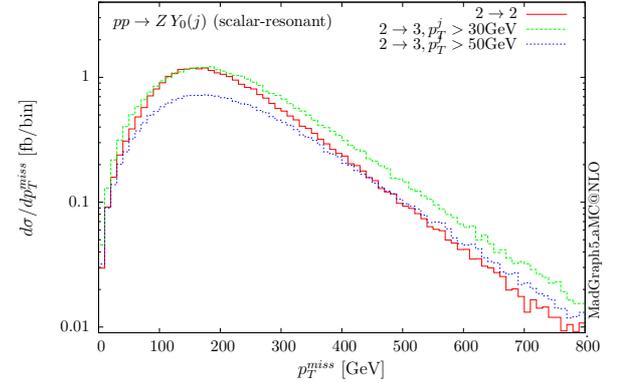}
\caption{Missing $p_T$ distribution for $ p  p \to Y_0 Z (j)$ for the scalar mediator of 200 GeV. The $p  p \to Y_0 Z j$ distributions are shown for two cuts on the transverse momentum of the jet.}
\label{Zptj}
\end{figure}

For this kind of process, it is important to investigate the importance of additional QCD radiation, similarly to what we observe in the SM for $gg\to ZH$ \cite{Hespel:2015zea}.
We illustrate this effect in Fig.~\ref{Zptj} by showing in the `resonant' case (200 GeV scalar mediator) the missing transverse momentum for the  zero and one-jet multiplicities.
 We see that these contributions are not suppressed compared to the 0-jet ones. In the SM this effect, i.e. the importance of the 1-jet contribution, is even more pronounced at high $p_T$ as the 0-jet amplitude is extremely suppressed by the cancellation between box and triangle diagrams. In any case,  it might be important experimentally to perform inclusive searches rather than applying a jet veto, as such a cut will eliminate these enhanced 1-jet contributions. A more accurate prediction of the shapes can be provided by LO merging and matching to the PS, a procedure that is automatic in \mgamc.

\subsection{mono-Higgs}
Studies of mono-Higgs production related to dark matter searches, have already been presented in the literature. In the scenario studied in \cite{Petrov:2013nia,Berlin:2014cfa}, the dark matter particles only couple to the Higgs and therefore the only production mode is in association with a Higgs boson. In addition to the mono-Higgs production through the coupling of the mediator to the Higgs, the gluon fusion contribution to the mono-Higgs production in the infinite top-mass limit has been studied in  \cite{Carpenter:2013xra}.

The gluon fusion contribution with the exact top-mass dependence can be obtained with our implementation of the simplified model. Mono-Higgs production in this simplified model is not possible with a vector mediator, similarly to $gg \to H\gamma$ production in the SM due to charge conjugation invariance.\footnote{We note here that these selection rules apply only to the $2\to 2$ scattering amplitudes. The amplitudes for these processes are non-zero when additional QCD radiation is allowed.} Therefore for this process we consider only scalar, pseudoscalar and axial vector mediators.  The corresponding  total cross-sections are given in Table \ref{tab:monoH}. The scalar cross-sections are of the same order but a bit larger than the SM $HH$ production  \cite{Frederix:2014hta} due to the lack of the destructive interference with the triangle with the trilinear Higgs coupling.

\begin{table*}[t]
\renewcommand{\arraystretch}{1.3}
\begin{center}
    \begin{tabular}{ l ? c | c | c  }
        \hline \hline
 Benchmark      & scalar         & pseudoscalar         & axial-vector\\        \hline
Resonant & 6.98\, \cdot $10^{-2} \, {}^{+34\%}_{-24\%}\,{}^{+1.0\%}_{-1.2\%}$ & 0.139\, ${}^{+33\%}_{-23\%}\,{}^{+1.0\%}_{-1.2\%}$ & 2.81\cdot $10^{-2} \, {}^{+36\%}_{-25\%}\,{}^{+1.3\%}_{-1.4\%}$  \\ 
Heavy mediator & 9.31\, \cdot $10^{-5} \, {}^{+41\%}_{-27\%}\,{}^{+2.1\%}_{-2.1\%}$ & 5.79\, \cdot $10^{-5} \, {}^{+40\%}_{-27\%}\,{}^{+1.9\%}_{-1.9\%}$  & 3.01\cdot $10^{-5} \, {}^{+41\%}_{-27\%}\,{}^{+2.1\%}_{-2.1\%}$  \\
Heavy DM & 1.28\, \cdot $10^{-7} \, {}^{+43\%}_{-28\%}\,{}^{+3.0\%}_{-2.9\%} $& 2.44\, \cdot $10^{-7} \, {}^{+42\%}_{-28\%}\,{}^{+2.6\%}_{-2.6\%}$ & 2.07\cdot $10^{-5} \, {}^{+43\%}_{-28\%}\,{}^{+2.9\%}_{-2.9\%}$ \\
\hline \hline
\end{tabular}
 \caption{\label{tab:monoH} Cross sections (in pb) for gluon induced mono-Higgs production
   at the LHC at $\sqrt{s} = 13$~TeV for the three mass benchmarks. A technical cut of 2~GeV has been set on the transverse momentum of all final heavy states but no Higgs branching ratios are included. }  
\end{center} 
\end{table*}

\begin{figure}[h!]
\centering
\includegraphics[trim=1.2cm 0 0 0 cm,scale=0.56]{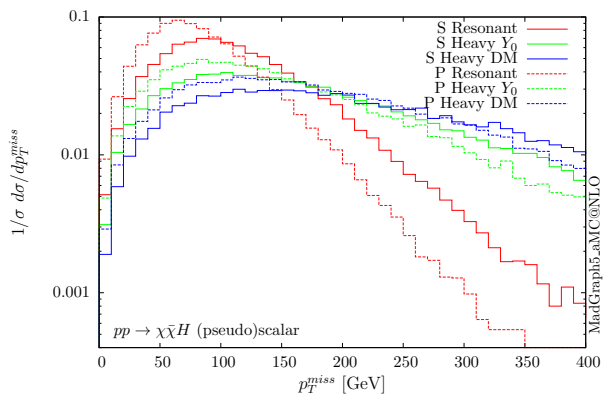}
\caption{Missing $p_T$ distribution for $ p  p \to \chi \bar{\chi} H$ for the scalar (S) and pseudoscalar (P) mediators.}
\label{Hpt_scalar}
\end{figure}

\begin{figure}[h!]
\centering
\includegraphics[trim=1.2cm 0 0 0 cm,scale=0.56]{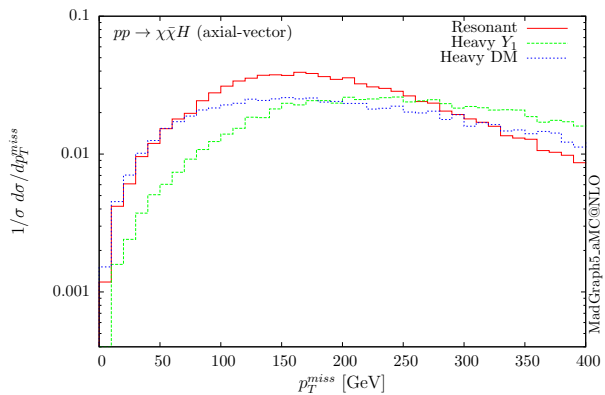}
\caption{Missing $p_T$ distribution for $ p  p \to \chi \bar{\chi} H$ for the axial-vector mediator. }
\label{Hpt_axial}
\end{figure}

The parton-level results for the normalised distributions of the missing $p_T$ for the benchmarks above are shown in Fig. \ref{Hpt_scalar} and \ref{Hpt_axial}, for the scalar, pseudoscalar and axial-vector mediators respectively. We observe different distributions for the scalar scenarios compared to the pseudoscalar ones. Indeed the form-factors describing the scattering amplitudes  are different depending on the parity of the scalar. In the infinite top mass limits, these differ exactly by a factor of 2/3.

\subsection{mono-photon}
\begin{figure}[h!]
\includegraphics[trim=1.2cm 0 0 0 cm, scale=0.56]{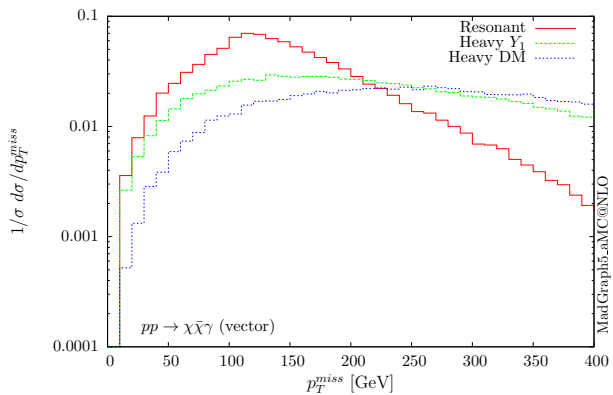}
\caption{Missing transverse momentum distribution for $ p  p \to \chi \bar{\chi} \gamma $ for the three mass benchmarks. The only mediator giving non zero results is the vector one.  }
\label{Apt}
\end{figure}

For the mono-photon case the only mediator which contributes is the vector one. The contributions of a scalar, pseudoscalar and axial-vector vanish due to charge conjugation invariance. The cross-sections at 13~TeV for the three benchmarks are shown in Table \ref{tab:photon}, while the corresponding distributions for the missing transverse energy are shown in Fig.~\ref{Apt}. Only the resonant scenario gives a non-negligible cross-section for unit couplings, implying that the mono-photon process in this simplified DM model can be accessible experimentally only for limited corners of the mass parameter space.

\begin{table*}[t]
\renewcommand{\arraystretch}{1.3}
\begin{center}
    \begin{tabular}{    r ? c   }
        \hline \hline
 Benchmark      & vector \\        \hline
Resonant & 3.18\, \cdot $10^{-2} \, {}^{+34\%}_{-24\%}\,{}^{+1.1\%}_{-1.2\%}$   \\ 
Heavy mediator & 1.98\, \cdot $10^{-5} \, {}^{+41\%}_{-27\%}\,{}^{+2.1\%}_{-2.1\%} $   \\
Heavy DM & 1.17\, \cdot $10^{-7} \, {}^{+43\%}_{-28\%}\,{}^{+3.0\%}_{-2.9\%} $  \\
\hline \hline
\end{tabular}
 \caption{\label{tab:photon} Cross sections (in pb) for gluon induced mono-photon production
   at the LHC at $\sqrt{s} = 13$~TeV for the three mass benchmarks. A 10~GeV cut is applied on the transverse momentum of the final state photon and $|\eta_{\gamma}|<2.5$.}  
\end{center} 
\end{table*}

\section{Conclusions}
\label{conclusions}
We have presented results of the implementation of a simplified dark matter model within \mgamc, involving a spin-0 or spin-1 mediator, coupling preferentially to the top quark. For this model the production of DM particles proceeds through gluon fusion loops. We have considered the production in association with QCD jets, leading to the jets plus missing transverse momentum signature at the LHC.  The results obtained depend strongly on the mass of the mediator and dark matter particles, leading to sharper or broader features in the differential distributions. To provide a reliable description of the distribution shapes we have employed matrix-element-parton shower merging and matching, and presented results for four mediator types: scalar, pseudoscalar, vector and axial-vector. Even though we have not considered them here, processes with mediators with mixed parity (scalar/pseudoscalar and vector/axial-vector) can be also simulated. 

In addition to the jets associated production, we also considered the production of DM in association with a $Z$, Higgs and photon. A subset of these production modes are forbidden by conservation laws, depending on the mediator type. Therefore it is important to complement searches with jets with searches for mono-X,  as any signal in these channels can straightforwardly provide information on the nature of the mediator.  

Our implementation is completely general and public. Even though in this study, we presented results at a rather inclusive level, the interested reader can use the implementation to obtain results with the appropriate cuts, mimicking the corresponding experimental searches. These results combined with the experimental measurements can be used to extract limits on the various model parameters, also taking into account constraints from direct and indirect detection experiments. 

\section*{Acknowledgements}
We would like to thank F. Maltoni for his continuous support during this project, V. Hirschi for his help on the loop induced part of \mgamc, the CP3 IT team for their constant support and K. Mawatari for useful discussions.
O.M. is a Durham International Junior Research Fellow.
This work is supported in part by the IISN  `{\sc\small MadGraph}' convention 4.4511.10, by the Belgian
Federal Science Policy Office through the Interuniversity Attraction Pole P7/37, by the European Union as part of the FP7 Marie Curie Initial Training Network MCnetITN (PITN-GA-2012-315877), and by the ERC grant 291377
`LHCtheory: Theoretical predictions and analyses of LHC physics:
advancing the precision frontier'.

\bibliographystyle{epjc}
\bibliography{DMbib}

\begin{thebibliography}{10}
\providecommand{\url}[1]{\texttt{#1}}
\providecommand{\urlprefix}{URL }
\providecommand{\eprint}[2][]{\url{#2}}

\bibitem{Carpenter:2013xra}
L.~Carpenter, A.~DiFranzo, M.~Mulhearn, et~al., Phys.Rev. \textbf{D89}, 7,
  075017 (2014), \eprint{1312.2592}

\bibitem{Abdallah:2014hon}
J.~Abdallah, A.~Ashkenazi, A.~Boveia, et~al.  (2014), \eprint{1409.2893}

\bibitem{Haisch:2015ioa}
U.~Haisch, E.~Re  (2015), \eprint{1503.00691}

\bibitem{Mao:2014rga}
M.~Song, G.~Li, W.-G. Ma, et~al., JHEP \textbf{1409}, 069 (2014),
  \eprint{1403.2142}

\bibitem{Fox:2012ru}
P.~J. Fox, C.~Williams, Phys.Rev. \textbf{D87}, 5, 054030 (2013),
  \eprint{1211.6390}

\bibitem{Wang:2011sx}
J.~Wang, C.~S. Li, D.~Y. Shao, et~al., Phys.Rev. \textbf{D84}, 075011 (2011),
  \eprint{1107.2048}

\bibitem{Huang:2012hs}
F.~P. Huang, C.~S. Li, J.~Wang, et~al., Phys.Rev. \textbf{D87}, 094018 (2013),
  \eprint{1210.0195}

\bibitem{Abdallah:2015ter}
J.~Abdallah, H.~Araujo, A.~Arbey, et~al.  (2015), \eprint{1506.03116}

\bibitem{Abercrombie:2015wmb}
D.~Abercrombie, et~al.  (2015), \eprint{1507.00966}

\bibitem{Frandsen:2012db}
M.~T. Frandsen, U.~Haisch, F.~Kahlhoefer, et~al., JCAP \textbf{1210}, 033
  (2012), \eprint{1207.3971}

\bibitem{Haisch:2013fla}
U.~Haisch, A.~Hibbs, E.~Re, Phys.Rev. \textbf{D89}, 3, 034009 (2014),
  \eprint{1311.7131}

\bibitem{Buckley:2014fba}
M.~R. Buckley, D.~Feld, D.~Goncalves, Phys.Rev. \textbf{D91}, 1, 015017 (2015),
  \eprint{1410.6497}

\bibitem{Harris:2014hga}
P.~Harris, V.~V. Khoze, M.~Spannowsky, et~al., Phys.Rev. \textbf{D91}, 5,
  055009 (2015), \eprint{1411.0535}

\bibitem{Alwall:2014hca}
J.~Alwall, R.~Frederix, S.~Frixione, et~al., JHEP \textbf{1407}, 079 (2014),
  \eprint{1405.0301}

\bibitem{Hirschi:2015iia}
V.~Hirschi, O.~Mattelaer  (2015), \eprint{1507.00020}

\bibitem{Hirschi:2011pa}
V.~Hirschi, et~al., JHEP \textbf{05}, 044 (2011), \eprint{1103.0621}

\bibitem{Ossola:2006us}
G.~Ossola, C.~G. Papadopoulos, R.~Pittau, Nucl.Phys. \textbf{B763}, 147 (2007),
  \eprint{hep-ph/0609007}

\bibitem{Ossola:2007ax}
G.~Ossola, C.~G. Papadopoulos, R.~Pittau, JHEP \textbf{0803}, 042 (2008),
  \eprint{0711.3596}

\bibitem{Mawatari}
M.~backovic, et~al., to appear  (2015)

\bibitem{Alloul:2013bka}
A.~Alloul, N.~D. Christensen, C.~Degrande, et~al., Comput.Phys.Commun.
  \textbf{185}, 2250 (2014), \eprint{1310.1921}

\bibitem{Degrande:2011ua}
C.~Degrande, C.~Duhr, B.~Fuks, et~al., Comput.Phys.Commun. \textbf{183}, 1201
  (2012), \eprint{1108.2040}

\bibitem{feynrules}
http://feynrules.irmp.ucl.ac.be/wiki/DMsimp

\bibitem{Alwall:2007fs}
J.~Alwall, et~al., Eur. Phys. J. \textbf{C53}, 473 (2008), \eprint{0706.2569}

\bibitem{Alwall:2011cy}
J.~Alwall, Q.~Li, F.~Maltoni, Phys.Rev. \textbf{D85}, 014031 (2012),
  \eprint{1110.1728}

\bibitem{Hespel:2015zea}
B.~Hespel, F.~Maltoni, E.~Vryonidou, JHEP \textbf{1506}, 065 (2015),
  \eprint{1503.01656}

\bibitem{Sjostrand:2006za}
T.~Sjostrand, S.~Mrenna, P.~Z. Skands, JHEP \textbf{05}, 026 (2006),
  \eprint{hep-ph/0603175}

\bibitem{Alwall:2008qv}
J.~Alwall, S.~de~Visscher, F.~Maltoni, JHEP \textbf{0902}, 017 (2009),
  \eprint{0810.5350}

\bibitem{Sjostrand:2007gs}
T.~Sjostrand, S.~Mrenna, P.~Z. Skands, Comput.Phys.Commun. \textbf{178}, 852
  (2008), \eprint{0710.3820}

\bibitem{Sjostrand:2014zea}
T.~Sjostrand, S.~Ask, J.~R. Christiansen, et~al.  (2014), \eprint{1410.3012}

\bibitem{Conte:2012fm}
E.~Conte, B.~Fuks, G.~Serret, Comput.Phys.Commun. \textbf{184}, 222 (2013),
  \eprint{1206.1599}

\bibitem{Conte:2014zja}
E.~Conte, B.~Dumont, B.~Fuks, et~al., Eur.Phys.J. \textbf{C74}, 10, 3103
  (2014), \eprint{1405.3982}

\bibitem{Cacciari:2011ma}
M.~Cacciari, G.~P. Salam, G.~Soyez, Eur.Phys.J. \textbf{C72}, 1896 (2012),
  \eprint{1111.6097}

\bibitem{Cacciari:2008gp}
M.~Cacciari, G.~P. Salam, G.~Soyez, JHEP \textbf{0804}, 063 (2008),
  \eprint{0802.1189}

\bibitem{Martin:2009iq}
A.~Martin, W.~Stirling, R.~Thorne, et~al., Eur.Phys.J. \textbf{C63}, 189
  (2009), \eprint{0901.0002}

\bibitem{SysCalc}
V.~Hirschi, A.~Kalogeropoulos, O.~Mattelaer, et~al., in preparation  (2015)

\bibitem{Alwall:2014bza}
J.~Alwall, C.~Duhr, B.~Fuks, et~al.  (2014), \eprint{1402.1178}

\bibitem{Demartin:2014fia}
F.~Demartin, F.~Maltoni, K.~Mawatari, et~al., Eur.Phys.J. \textbf{C74}, 9, 3065
  (2014), \eprint{1407.5089}

\bibitem{Dawson:1998py}
S.~Dawson, S.~Dittmaier, M.~Spira, Phys. Rev. \textbf{D58}, 115012 (1998),
  \eprint{hep-ph/9805244}

\bibitem{Landau:1948kw}
L.~D. Landau, Dokl. Akad. Nauk Ser. Fiz. \textbf{60}, 207 (1948)

\bibitem{Yang:1950rg}
C.-N. Yang, Phys. Rev. \textbf{77}, 242 (1950)

\bibitem{Glover:1988rg}
E.~W.~N. Glover, J.~J. van~der Bij, Nucl. Phys. \textbf{B321}, 561 (1989)

\bibitem{Petrov:2013nia}
A.~A. Petrov, W.~Shepherd, Phys.Lett. \textbf{B730}, 178 (2014),
  \eprint{1311.1511}

\bibitem{Berlin:2014cfa}
A.~Berlin, T.~Lin, L.-T. Wang, JHEP \textbf{1406}, 078 (2014),
  \eprint{1402.7074}

\bibitem{Frederix:2014hta}
R.~Frederix, S.~Frixione, V.~Hirschi, et~al., Phys.Lett. \textbf{B732}, 142
  (2014), \eprint{1401.7340}

\end{thebibliography}
\end{document}